\documentclass[fleqn,usenatbib]{mnras}

\usepackage{newtxtext,newtxmath}
\usepackage[T1]{fontenc}
\usepackage{ae,aecompl}

\usepackage{graphicx}	% Including figure files
\usepackage{amsmath}	% Advanced maths commands
\usepackage{url}
\usepackage{siunitx}
\usepackage{breqn}
\usepackage[utf8]{inputenc}
\usepackage{makecell}
\usepackage{rotating}
\usepackage{pdflscape}
\usepackage{afterpage}
 \usepackage[version=4]{mhchem}

\newcommand\av{$A_{\rm V}$}

\newcommand\gtsim{\mathrel{\hbox{\rlap{\hbox{\lower4pt\hbox{$\sim$}}}\hbox{$>$}}}}
\newcommand\ha{\ce{H\alpha}}

\newcommand\loiii{$L_{\rm [O\,{\scriptscriptstyle III}]}$}
\newcommand\Lsun{L$_{\odot}$}
\newcommand\ltsim{\mathrel{\hbox{\rlap{\hbox{\lower4pt\hbox{$\sim$}}}\hbox{$<$}}}}
\newcommand\nh{$N_{\rm H}$}
\newcommand\nhlos{$N_{\rm H}^{\rm los}$}
\newcommand\nhperp{$N_{\rm H}^{\perp}$}
\newcommand\oii{[O\,{\sc ii}]}
\newcommand\oiii{[O\,{\sc iii}]}
\newcommand\lmir{$L_{12}$}

\newcommand\neowise{{NEOWISE}}
\newcommand\wise{{\em WISE}}
\newcommand\sdss{{SDSS}}
\newcommand\boss{{BOSS}}
\newcommand\spitzer{{\em Spitzer}}

\newcommand\chandra{{\em Chandra}}
\newcommand\xmm{{\em XMM-Newton}}

\newcommand\xrt{{\em Swift}-XRT}
\newcommand\bat{{\em Swift}-BAT}

\newcommand\nustar{{\em NuSTAR}}
\newcommand\hst{{\em HST}}

\newcommand\first{{FIRST}}
\newcommand\twomass{{2MASS}}

\newcommand\xspec{{\sc xspec}}
\newcommand\agnfitter{{\sc agnfitter}}

\DeclareSIUnit\angstrom{\text {Å}}
\title[MIR quasars lacking narrow optical emission lines]{A population of Optically Quiescent Quasars from \wise\ and \sdss}
\author[Greenwell et al.]{Claire Greenwell,$^{1,3}$\thanks{E-mail: claire.l.greenwell@durham.ac.uk}
Poshak Gandhi,$^{1}$
Daniel Stern,$^{4}$
George Lansbury,$^{2}$
\newauthor
Vincenzo Mainieri,$^{2}$
Peter Boorman,$^{5,6,1}$
Yoshiki Toba$^{7,8,9}$
\\
$^{1}$School of Physics \& Astronomy, University of Southampton, Highfield, Southampton SO17 1BJ, UK.\\
$^{2}$European Southern Observatory, Karl-Schwarzschild-Strasse 2, D-85748
Garching, Germany.\\
$^{3}$Centre for Extragalactic Astronomy, Department of Physics, Durham University, Durham, DH1 3LE, UK.\\
$^{4}$Jet Propulsion Laboratory, California Institute of Technology, 4800 Oak Grove Drive, Mail Stop 169-221, Pasadena, CA 91109, USA.\\
$^{5}$Cahill Center for Astrophysics, California Institute of Technology, 1216 East California Boulevard, Pasadena, CA 91125, USA.\\
$^{6}$Astronomical Institute, Academy of Sciences, Boční II 1401, CZ-14131 Prague, Czechia.\\
$^{7}$National Astronomical Observatory of Japan, 2-21-1 Osawa, Mitaka, Tokyo 181-8588, Japan.\\
$^{8}$Academia Sinica Institute of Astronomy and Astrophysics, 11F of Astronomy-Mathematics Building, AS/NTU, No.1, Section 4, Roosevelt Road, Taipei 10617, Taiwan.\\
$^{9}$Research Center for Space and Cosmic Evolution, Ehime University, 2-5 Bunkyo-cho, Matsuyama, Ehime 790-8577, Japan.
}
\date{Accepted XXX. Received YYY; in original form ZZZ}

\pubyear{2023}

\hypersetup{draft}

\begin{document}
\label{firstpage}
\pagerange{\pageref{firstpage}--\pageref{lastpage}}
\maketitle

\begin{abstract}
% \textcolor{red}{250 word limit.}
The growth of active galactic nuclei (AGN) occurs under some form of obscuration in a large fraction of the population. The difficulty in constraining this population leads to high uncertainties in cosmic X-ray background and galaxy evolution models. Using an \sdss-\wise\ cross-match, we target infrared luminous AGN ($W1-W2$\,>\,0.8, and monochromatic rest-frame luminosity above $\lambda L_{\lambda}$(\SI{12}{\micro\metre})\,$\approx$\,3\,$\times$\,10$^{44}$\,erg\,s$^{-1}$), but with passive galaxy-like optical spectra (Optically Quiescent Quasars; OQQs). We find 47 objects that show no significant \oiii$\lambda$5007 emission, a typically strong AGN optical emission line. As a comparison sample, we examine \sdss-selected Type 2 quasars (QSO2s), which show a significant \oiii$\lambda$5007 line by definition. We find a 1:16 ratio of OQQs compared to QSO2s, suggesting that the OQQ duty cycle is likely much shorter than that of QSO2s (though selection biases are not fully quantified). We consider observed properties in comparison with other galaxy types, and examine them for consistency with theories on their intrinsic nature: chiefly (a) a high covering factor for surrounding obscuring matter, preventing the detection of high-ionisation emission lines -- `cocooned AGN'; or (b) ionised gas being absent on the kpc scales of the Narrow Line Region (NLR), perhaps due to a `switching on' or `young' AGN. OQQs do not obviously fit the standard paradigm for merger-driven AGN and host galaxy evolution, implying we may be missing part of the flow of AGN evolution.

\end{abstract}


%%%%%%%%%%%%%%%%%%%%%%%%%%%%%%%%%%%%%%%%%%%%%%%%%%

%%%%%%%%%%%%%%%%% BODY OF PAPER %%%%%%%%%%%%%%%%%%

\section{Introduction}

The census of active galactic nuclei (AGN) is, at present, highly incomplete. Dusty gas that feeds supermassive black hole growth can obscure the nucleus, resulting in an attenuation of AGN signatures along the line-of-sight (l.o.s.). This is not a minor effect: the {\em majority} of AGN are affected by obscuration. There are many lines of evidence pointing to this. Studies of the cosmic X-ray background radiation require obscuration by neutral gas with column densities (\nh) exceeding 10$^{22}$\,cm$^{-2}$ with between 2--5 as many times obscured AGN as unobscured ones \citep[e.g. ][]{setti_active_1989, fabian_mass_1999, gandhi_x-ray_2003, gilli_synthesis_2007, treister_space_2009, ueda_toward_2014,ananna_accretion_2019}. In the optical, obscuration can successfully explain the variety of observed AGN classes. Optical Type 1 AGN show broad emission lines with widths of several thousand km\,s$^{-1}$ arising on scales of $\sim$\,30\,$L_{5100}^{0.7}$\,light-days, where $L_{5100}$ is the continuum rest-frame luminosity at 5100\,\AA\ in units of 10$^{44}$\,erg\,s$^{-1}$ \citep[e.g. ][]{kaspi_reverberation_2000}. Observing these close nuclear scales requires an extinction-free l.o.s.. Type 1 and Type 2 AGN show narrower emission lines with widths of a few hundred km\,s$^{-1}$, which arise on scales of tens to thousands of pc. This difference in appearance can naturally be explained by an anisotropic distribution of dust which obscures the close-in broad lines, but not the larger scale narrower ones. In this way, the zoo of AGN classes can be unified. This model can also explain observed polarization fractions of broad emission lines \citep[e.g. ][]{antonucci_spectropolarimetry_1985, agn_peterson}. In the radio, the radio-loud\footnote{Defined as galaxies that have significant emission at radio wavelengths; radio (5GHz) to optical (\textit{B}-band) ratio $\gtsim$10 \citep{kellermann_vla_1989}} subset of obscured quasars (\lq radio galaxies\rq) was the first significant population of powerful and heavily absorbed AGN to be followed up in detail \citep[e.g. ][]{mccarthy_high_1993, miley_distant_2008, toba_wide_2019}; many X-ray studies have now shown them to be strongly obscured, on average \citep[e.g. ][]{gandhi_4c_2006, tozzi_vla_2009, wilkes_revealing_2013}.

The column density and geometry of obscuring matter are expected to naturally evolve as AGN and their host galaxies grow, and many models posit that the bulk of supermassive black hole growth occurs in highly obscured phases \citep[e.g. ][]{fabian_obscured_1999, dimatteo_energy_2005, hopkins_unified_2006}. AGN are also known to appear to change classification over time, from Type 1 to 2 and vice versa \citep{yang_discovery_2018}. These objects, referred to as `Changing Look' AGN, show changes in emission line and continuum flux over timescales up to a few years. The physical mechanisms behind these changes are not well understood: the two main theories are variation in the line-of-sight obscuration (e.g. a clumpy torus; \citealp{elitzur_on_2012}), a change in the accretion rate \citep[e.g.][]{sheng_mid-infrared_2017}, or thermal changes in the inner accretion disk \citep[e.g.][]{stern_mid-ir_2018}. Any unusual AGN appearance must be considered in this context - lack of emission lines could be a transitional state of changing obscuration, or a change in intrinsic line production. Objects in the process of `switching on' are a rare, brief chapter in the growth of AGN, an important but not well understood period. Sources with extreme covering factors approaching unity could also probe a unique phase in AGN evolution, indicating either strong growth rates with plenty of available circumnuclear matter for accretion or perhaps sky covering as a result of merger-driven turbulence. Several studies have linked galaxy mergers with higher rates of obscured AGN in MIR selected samples \citep[e.g. ][]{glikman_first-2mass_2012, weston_incidence_2017, satyapal_buried_2017}. Although Seyfert galaxies with high covering factors approaching unity have been inferred in detailed individual studies \citep{ramos_almeida_infrared_2009}, the fraction of highly covered AGN at {\em high} power is expected to be small (e.g. \citealp{toba_luminosity_2014,stalevski_dust_2016}, although the dependence on luminosity is weak and not without counter-evidence - e.g. \citealp{netzer_star_2016}). Theoretically, accretion from large scales is not expected to be isotropic and is likely to be mediated via disks, warps and other instabilities \citep[e.g. ][]{hoenig_redefining_2019}.

The infrared regime is particularly effective for studies of AGN dust covering factors, and for studies of AGN with absorbed optical signatures. This is because dust serves as a bolometer, absorbing and reprocessing the AGN power to the mid-infrared (MIR), providing a probe of the obscuring material in {\em emission}, as opposed to the {\em absorption} pathway provided in optical and X-ray studies. High angular resolution multiwavelength studies suggest that the MIR emission is effectively (to within a factor of a few) isotropic \citep{gandhi_resolving_2009, levenson_isotropic_2009, asmus_subarcsecond_2015, stalevski_dust_2016}. Covering fractions derived from MIR AGN number counts and modelling of individual spectral energy distributions may also be a function of luminosity \citep{maiolino_dust_2007, alonso-herrero_torus_2011, assef_mid-infrared_2013, toba_9_2013, toba_luminosity_2014, toba_how_2021}, though this remains controversial \citep{roseboom_ir-derived_2013, lawrence_misaligned_2010, assef_half_2015}. 

One limitation of most works on covering fraction to-date is the requirement for the presence of AGN emission lines (typically forbidden lines from the Narrow Line Region; NLR) in the optical or near-infrared, used to confirm the presence of an AGN and/or redshift identification. If AGN emission lines are observed, this implies that some intrinsic power must be escaping the AGN environment and the source cannot be fully covered. The presence of dust in AGN NLRs has been known for some time \citep[e.g.,][]{netzer_dust_1993, haas_spitzer_2005, netzer_correlation_2006} and can attenuate forbidden line fluxes, but selecting {\em fully} covered AGN is difficult, given the obvious observational biases against identifying such a population. Thus, there have only been a limited number of studies of AGN in  the high covering factor regime \citep[e.g.,][]{gandhi_very_2002, imanishi_spitzer_2007}. The \oiii$\lambda$5007 line is one of the strongest optical AGN emission lines and can be used as a proxy for bolometric luminosity \citep[e.g.,][]{heckman_present-day_2004}.  It arises in the NLR, likely as a result of photoionization by AGN radiation. This places the line origin beyond the putative classical torus of unification models, so it should not be affected by l.o.s. nuclear reddening \citep[although there is some effect, e.g. on the bolometric correction;][]{lamastra_bolometric_2009}. Narrow line emission, and in particular \oiii, is assumed to be present in the majority of AGN and is often used for selection; e.g., BPT diagrams \citep{kewley_host_2006} and Type 2 quasars \citep{zakamska_candidate_2003}. Sources lacking this line are unusual among AGN catalogues, and by selecting based on this absence we can target objects either (a) with unstable large scale obscuration preventing transmission of \oiii, or (b) in which the physical conditions result in no line formation in the first place; for example if the AGN is recently switched on, and the ionising radiation has not yet reached NLR scales. 

Recent large surveys now enable studies searching for elusive AGN subtypes to be carried out. Here, we go beyond previous works in selecting candidate mid-infrared (MIR) quasars that show no optical signatures. For this, we use the latest all-sky MIR survey by the \textit{Wide-field Infrared Survey Explorer} \citep[\wise;][]{wright_wide-field_2010} mission, combined with optical spectroscopy from the Sloan Digital Sky Survey (\sdss) to select MIR--luminous sources with AGN--like MIR colours but with early-type galaxy optical spectra. In this way, we avoid star-forming galaxies and associated degeneracies in separating such systems from AGN \citep[e.g. ][]{trouille_optx_2010}. At lower luminosity, spectral dilution of AGN emission by the host galaxy cannot be neglected \citep[e.g. ][]{moran_hidden_2002, comastri_nature_2002, cocchia_hellas2xmm_2007, civano_hellas2xmm_2007, caccianiga_elusive_2007}. At high luminosity, it becomes increasingly difficult for the host galaxy to dilute the AGN, so MIR quasar studies offer a clean probe of nuclear activity.

The structure of this paper is as follows. We describe the data used and our sample selection in Sections \ref{sec:data} and \ref{sec:sampleselection}. Results follow in Section \ref{sec:results}. A detailed discussion of implications, caveats, and comparisons with other source classes can be found in Section \ref{sec:discussion}. An Appendix includes information on individual sources, the optical spectra and broadband SEDs. We assume a flat cosmology with $H_{\rm 0}$\,=\,67.4\,km\,s$^{-1}$\,Mpc$^{-1}$ and $\Omega_\Lambda$\,=\,0.685 \citep{planck_collaboration_planck_2018}.

\section{Data} \label{sec:data}

\subsection{\wise}

The \wise\ satellite has carried out a highly-sensitive all-sky survey in four bands ($W1, W2, W$3 and $W4$, centred on wavelengths of 3.4, 4.6, 12 and 22 \SI{}{\micro\metre}, respectively). The effective angular resolution corresponds to a Gaussian with full-width-at-half-maximum \SI{6}{\arcsecond} in $W1-W3$, and \SI{12}{\arcsecond} in $W4$.  The \sdss\ fifteenth Data Release \citep{aguado_fifteenth_2019} includes pre-calculated astrometric cross matches with the earlier \wise\ data release (\wise-AllSky\footnote{\url{http://wise2.ipac.caltech.edu/docs/release/allsky/}}). The AllWISE release\footnote{\url{http://wise2.ipac.caltech.edu/docs/release/allwise/}} includes data from both the cryogenic and post-cryogenic phases of the mission, and therefore contains better quality data, with better photometric sensivitity, particularly in the $W1$ and $W2$ bands. The cross matching with \sdss\ and \wise-AllSky was used, and the \wise\ magnitudes updated with AllWISE values.

We choose to use the AllWISE data rather than the more recent CatWISE \citep{marocco_catwise2020_2021} because the $W3$ and $W4$ measurements are vital to the selection process, and these two bands were only in operation in the earlier cryogenic years of \wise. Although CatWISE may have provided more accurate photometry, (a) we are only looking at bright sources, which should be present in the earlier catalogs, and (b) in case of variation in source output it is best to use measurements from all four bands taken in the same time period.

\subsection{\sdss}

Data from the \sdss\ data release 15 (DR15; \citealt{aguado_fifteenth_2019}), including Baryon Oscillation Spectroscopic Survey (\boss) spectra \citep{dawson_baryon_2013}, were used for cross-matching with the \wise\ catalog. 

The relevant tables are:

\begin{itemize}
    \item {\tt SpecObjAll}: the base \sdss\ table for spectroscopic observations, containing all measured spectra, including potentially bad or duplicate data. Contains 4,851,200 sources.
    \item {\tt SpecObj}: the sub table of {\tt SpecObjAll}, containing only the clean data, filtered for duplicates. Contains 4,311,571 sources.
    \item {\tt PhotoObjAll}: the base \sdss\ table for all photometric observations. Contains 1,231,051,050 measurements.
    \item {\tt PhotoObj}: the subset of {\tt PhotoObjAll} containing only primary and secondary objects (i.e. not a family object, outside the chunk boundary, or within a hole\footnote{\url{https://www.sdss.org/dr15/algorithms/masks/}}).
    Contains 794,328,715 measurements.
    \item {\tt wise\_allsky}: the \wise\ catalog from the \wise-AllSky data release. Contains 563,921,584 sources.
    \item {\tt wise\_xmatch}: a joined table that contains pointers to \sdss\ and \wise-AllSky measurements, from astrometric cross-matches between the two. Contains 495,003,196 cross-matches.
\end{itemize}

\section{Sample selection}
\label{sec:sampleselection}

\begin{figure}
	\includegraphics[width=\columnwidth]{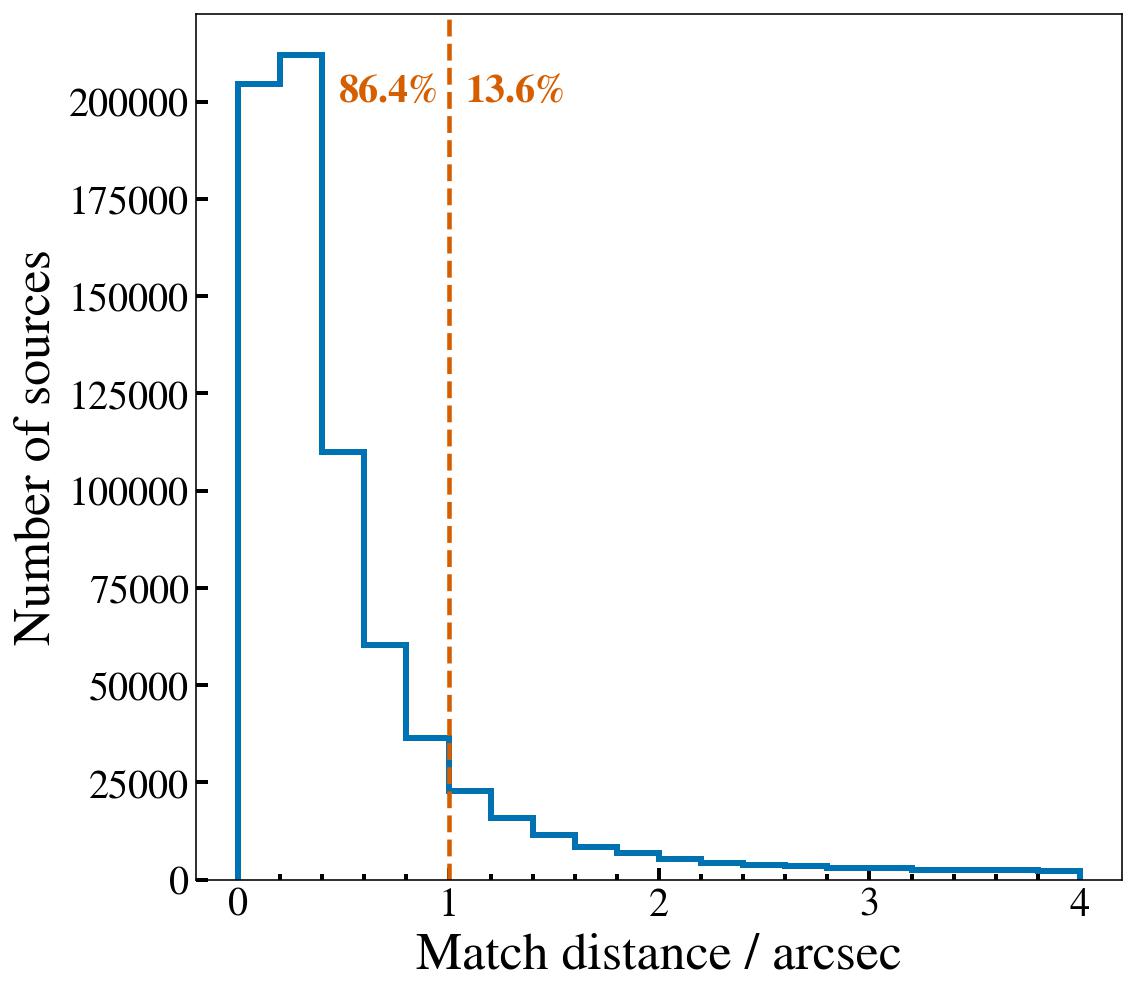}
    \caption{\sdss--\wise\ nearest counterpart cross-matching distance distribution. Numbers in orange indicate the fraction of sources in the accepted (left) and rejected (right) sections.}
    \label{fig:sep}
\end{figure}

The initial \sdss\ to \wise\ crossmatch (table {\tt wise\_xmatch}) uses a 4 arcsecond distance threshold. The nominal \wise\ positional accuracy is actually significantly better than this\footnote{\url{http://wise2.ipac.caltech.edu/docs/release/allwise/expsup/sec2_5.html}}. This is also seen in the distribution of associated counterpart distances shown in Fig.\,\ref{fig:sep}, which peaks at $<$\SI{0.2}{\arcsecond}. A few selection tests showed that an increasing fraction of ambiguous counterpart identification (e.g. two potential \sdss\ sources for a single \wise\ source) for distances of more than 1 arcsecond. In fact, the vast majority (86.4\%) of closest associations lie at distances of less than \SI{1}{\arcsecond}, so this is the threshold we selected to build our catalog. We have chosen to do cross-matching based on positional separation only rather than using a more statistical method \citep[e.g., \textsc{Nway},][]{salvato_finding_2018} due to the low incidence of multiple potential counterparts. The probability of each match being correct is high, based on low \wise\ positional uncertainties.

The initial set was based on the {\tt wise\_xmatch} catalog, with every spectral and photometric match joined (where one existed). The total number of objects in this table is 495,003,196\footnote{3252 of these have W1 magnitude=9999, i.e. are not usable.}. From these, we selected sources using the following criteria:

\begin{enumerate}

\item $\mathbf{W1-W2\ge 0.8}$: This colour threshold has been shown by \citet{stern_mid-infrared_2012} to be an efficient AGN selection criterion, yielding AGN samples with very high ($\sim$95\%) reliability and good ($\sim$80\%) completeness at high X-ray luminosities. AGN are expected to heat dust to temperatures of several hundred degrees K, approaching sublimation, resulting in an SED peaking at a few microns and producing a red $W1-W2$ colour. Whereas the colour cut alone can be contaminated by cool brown dwarfs, dust-reddened stars and star-forming galaxies \citep[e.g. ][]{stern_mid-infrared_2012, yan_characterizing_2013, hainline_spectroscopic_2014, assef_wise_2018}, our additional luminosity selection of only the most powerful sources is expected to weed out most such contaminants, preferentially selecting AGN alone (see step \ref{stepvi}). We will return to this point in Section \ref{sec:discussion}.

\item \textbf{\wise\ - \sdss\ cross match distance} - required to be $<$\SI{1}{\arcsecond} as discussed above.

\item \textbf{Reliable redshift:}\begin{itemize}
    \item {\tt z\_ERR/z\,$\le$\,0.01} - a fractional error of 1\,\% or less on the redshift, indicating an accurate value. This criterion was introduced because a good {\tt zWarning} flag does not necessarily imply a robust redshift measurement. Note that the median fractional $z$ uncertainty in our final sample is 0.00016, i.e. much smaller than the adopted threshold of 1\,\% which is designed to weed out the most obvious unreliable sources.
    \item {\tt zWarning} flag (\sdss) = 0, indicating high confidence in the redshift. Possible causes for a warning flag include poor or inconclusive fits, insufficient wavelength coverage, or problems with the instrumentation.
\end{itemize}

\item \textbf{Redshift limit:} $0 < z \le 0.83$ (for \sdss\ spectroscopy) or $0 < z \le 1.065$ (for \boss\ spectroscopy) - this ensures that all spectra cover the redshifted \oiii$\lambda$5007 line. For \sdss, this is identical to the redshift cut adopted by \citet{reyes_space_2008} for selecting \oiii--luminous Type 2 quasars, thus allowing us to contrast this class of objects with our targets. We additionally include spectra from newer survey programmes, and as these use a wider wavelength range spectrometer, we can include some further objects.

\item \textbf{Reliable \wise\ data:}\begin{itemize}
    \item \wise\ {\tt cc\_flags}\,=\,\lq 0000\rq. The four zeroes refer to the four \wise\ bands and indicate that the \wise\ data are not affected by any known artifacts that can cause confusion or contamination in any of the bands.
    \item \wise\ {\tt W?ph\_qual}\,$\ne$\,\lq U\rq\ AND {\tt W?ph\_qual}\,$\ne$\,\lq X\rq\ AND {\tt W?ph\_qual}\,$\ne$\,\lq Z\rq. This ensures a detection, reliable photometry, and measurable uncertainty in all \wise\ bands (with a signal:noise always above 2 and in the majority of sources above 3), allowing a robust measurement of the MIR source monochromatic fluxes (see next point) as well as MIR SEDs.
\end{itemize}

\item \textbf{IR luminosity:} \lmir $\,\ge 3\,\times\,10^{44}$\,erg\,s$^{-1}$, where \lmir\ is the $k$-corrected monochromatic \SI{12}{\micro\metre} luminosity ($\lambda L_{\lambda}$), computed by simple linear interpolation between log($\lambda L_{\lambda}(W3)$) and log($\lambda L_{\lambda}(W4)$) to rest-frame. For the MIR\,vs.\,X-ray relation of \citet{gandhi_resolving_2009}, this \lmir\ corresponds to an intrinsic 2-10 keV X-ray luminosity of $\ge$\,10$^{44}$\,erg\,s$^{-1}$, widely adopted as the threshold for selection of sources with quasar X-ray luminosities (e.g. \citealt{gandhi_powerful_2004, mainieri_black_2011, brusa_high-redshift_2009}). Furthermore, the \wise\ extragalactic source population above this \SI{12}{\micro\metre} luminosity threshold (corresponding to $\approx$\,10$^{11}$\,\textit{\Lsun}) comprises AGN almost exclusively \citep{donoso_origin_2012}.\label{stepvi}

\item \textbf{\sdss\ classification:} {\tt CLASS}\,=\,\lq{\tt GALAXY}\rq\ and {\tt SUBCLASS}\,=\,\lq\rq\ - the {\tt CLASS} selection removes sources with detected emission lines characteristic of QSOs (stars already having been culled by the redshift criteria).
The {\tt SUBCLASS} selection further removes sources with emission lines characteristic of Seyfert and LINERs, as well as star-forming and starburst systems\footnote{\url{https://www.sdss.org/dr15/spectro/catalogs/\#Objectinformation}}.

\item \textbf{Best available spectrum:} {\tt SciencePrimary}=1. This indicates that the spectra are considered by \sdss\ to be the best available for each object.

\item \textbf{No significant detection of \oiii$\lambda$5007}. The {\tt emissionLinesPort} table contains detailed flux fitting of a large number of \sdss\ spectra, but has the disadvantage of not having been applied to many newer spectra. In order to include the maximum number of possible spectra, we measured the significance of the \oiii$\lambda$5007 line flux directly, similarly to the method used to fit \oiii\ lines when selecting for QSO2s \citep{reyes_space_2008}: Gaussian curves were fit to rest frame positions of \oiii$\lambda$5007 in the spectral data from \sdss. The outcome of this trial was that a subset of the objects had no apparent \oiii\ emission line, and therefore these will comprise the final sample. A more detailed explanation of this trial is presented in Appendix \ref{app:oiiicheck}.

\item \textbf{Continuum SNR limit:} Estimate of SNR in the direct region around where an \oiii\ line would be detected. This is to remove any objects where the noise in the region of interest is overwhelming the measurement, leading to artificially high upper limits for the line flux. A more detailed explanation of this trial is presented in Appendix \ref{app:snrcheck}.

\item \label{step:vi} \textbf{Visual inspection of sources:} Automatic classification is not perfect, and a number of sources with e.g. obviously wrong redshift, classification, or visible emission lines were rejected in this final step.

\end{enumerate}

An overview of the selection steps can be seen in Table \ref{tab:selection}.

\subsection{Sample Contaminants}\label{sec:contaminants}

Although the selection steps are designed conservatively to produce a clean sample, we must consider the possibility of contaminant objects; i.e. sources that pass all the same criteria but are not predominantly AGN powered.
\begin{itemize}
    \item \textbf{Blazars} (as discussed further in Section~\ref{sec:blazars}) are highly luminous and are generally lacking strong emission lines. However, the shape of their optical spectrum is often very distinct, and any clear blazars were removed in step~\ref{step:vi}.
    \item \textbf{Luminous dusty star-forming galaxies} may reach luminosities comparable to the OQQ threshold in rare cases. To assess this probability, we compare with the catalogue produced by \citet{chang_stellar_2015}, selecting non-quiescent galaxies (using the threshold from \citealp{carnall_massive_2023}) and processing them through the OQQ MIR selection. We find that only $\sim$0.01\% would pass the colour and luminosity criteria. If we also include the \sdss\ spectroscopic requirement for a {\tt null} classification based on emission lines, no sources remain. This is logical, as any star formation intense enough to produce MIR emission on this level is likely to show strong emission lines on a galactic scale. The amount and distribution of dust that would be required to reduce the fluxes of these lines to the low levels detected (or upper limits of non-detections) would be highly unrealistic.
    \item \textbf{Compact Obscured Nuclei (CONs)} (as discussed further in Section~\ref{sec:ulirgs}) may represent a more likely class of contaminants, although still low probability overall. These sources contain a dense, IR-bright core that may be powered by compact star formation or an AGN, but are usually found in LIRGs, ULIRGs, and disturbed systems which typically show strong optical emission lines. CONs are currently thought to be rare, but both their intrinsic power source and true number counts are unknown. We should note that they would only be considered contaminants if they are powered by star-formation - if they contain a strong AGN and reside in quiescent galaxies they would be correctly included as OQQs.
\end{itemize}

\subsection{Comparison Population}\label{sec:qso2s}

As a comparative and complementary sample, we use a combination of:

\begin{itemize}
    \item The \sdss-selected sample of Type 2 quasars from \citet{reyes_space_2008}. These are sources that are high luminosity ($L_{\mathrm{bol}}>10^{45}$ erg s\textsuperscript{-1}), show a significant \oiii$\lambda$5007 line, and are optically obscured \citep{reyes_space_2008}. 
    \item A later, but similar, sample from \citet{yuan_spectroscopic_2016} selecting QSO2s from both \sdss\ and \boss\ spectroscopy.
\end{itemize}

\begin{landscape}
\begin{table}
    \centering
    \caption{Basic information about the 47 OQQs that passed the selection tests. A machine readable table can be found online. Column details: (1) short object name; (2),(3) sky position; (4) redshift; (5) rest frame \SI{12}{\micro\metre} luminosity interpolated from $W3$ and $W4$ luminosities; (6) rest frame 5100 \AA\ luminosity; (7) 4000 \AA\ break; (8), (9) \oiii, \ha\ luminosity upper limits in units of $L_\odot$; (10), (11) Minimum \av\ derived from different emission line luminosities; (12), (13) \nh, derived from \av.}
	\label{tab:allobjdetails}
	\resizebox{\columnwidth}{!}{
    \begin{tabular}{lrrrlllllllll}
\hline
 \thead{Name\\(1)}   &   \thead{RA (J2000)\\(2)} &   \thead{DEC (J2000)\\(3)} &   \thead{$z$\\(4)} & \thead{log $L_{12}$\\erg s$^{-1}$\\(5)}   & \thead{log $L_{5100}$\\erg s$^{-1}$\\(6)}   & \thead{D4000\\(7)}   & \thead{\oiii$\lambda$5007\\ $L_\odot$\\(8)}   & \thead{\ha$\lambda$6562\\ $L_\odot$ \\(9)}   & \thead{$A_V$\\(\oiii\ derived)\\(10)}   & \thead{$A_V$\\(\ha\ derived)\\(11)}   & \thead{log $N_H$ (cm$^{-2}$)\\(\oiii\ derived)\\(12)}   & \thead{log $N_H$ (cm$^{-2}$)\\(\ha\ derived)\\(13)}   \\
\hline
 OQQ J0002-0025      &                     0.592 &                     -0.432 &              0.371 & 44.81 $\pm$ 0.02                           & 44.00 $\pm$ 0.04                             & 1.32 $\pm$ 0.21      & $<$6.69                           & 8.15                            & 5.87                                    & --                                    & 22.91                                                  & --                                                   \\
 OQQ J0008+3144      &                     2.226 &                     31.749 &              0.600 & 44.71 $\pm$ 0.10                           & 44.03 $\pm$ 0.08                             & 1.71 $\pm$ 0.41      & --                                & --                              & --                                      & --                                    & --                                                     & --                                                   \\
 OQQ J0021+3515      &                     5.362 &                     35.264 &              0.618 & 45.27 $\pm$ 0.03                           & 44.00 $\pm$ 0.08                             & 1.51 $\pm$ 0.27      & $<$6.26                           & --                              & 8.15                                    & --                                    & 23.05                                                  & --                                                   \\
 OQQ J0059+2502      &                    14.757 &                     25.049 &              0.802 & 45.02 $\pm$ 0.17                           & 44.21 $\pm$ 0.11                             & 1.22 $\pm$ 0.21      & $<$6.00                           & --                              & 8.14                                    & --                                    & 23.05                                                  & --                                                   \\
 OQQ J0101+0731      &                    15.309 &                      7.518 &              0.563 & 44.68 $\pm$ 0.14                           & 44.40 $\pm$ 0.09                             & 1.36 $\pm$ 0.14      & $<$7.20                           & 7.98                            & 4.24                                    & --                                    & 22.77                                                  & --                                                   \\
 OQQ J0103-0349      &                    15.792 &                     -3.830 &              0.507 & 44.52 $\pm$ 0.11                           & 43.69 $\pm$ 0.11                             & 1.28 $\pm$ 0.28      & $<$6.67                           & 8.06                            & 5.17                                    & --                                    & 22.85                                                  & --                                                   \\
 OQQ J0109+0103      &                    17.379 &                      1.063 &              0.784 & 45.24 $\pm$ 0.13                           & 44.47 $\pm$ 0.05                             & 1.36 $\pm$ 0.08      & $<$6.26                           & --                              & 8.06                                    & --                                    & 23.05                                                  & --                                                   \\
 OQQ J0114+2002      &                    18.622 &                     20.047 &              0.589 & 44.61 $\pm$ 0.14                           & 44.13 $\pm$ 0.08                             & 1.59 $\pm$ 0.25      & $<$4.74                           & --                              & 10.22                                   & --                                    & 23.15                                                  & --                                                   \\
 OQQ J0141+1050      &                    25.294 &                     10.835 &              0.580 & 44.65 $\pm$ 0.11                           & 44.01 $\pm$ 0.09                             & 1.52 $\pm$ 0.29      & $<$6.78                           & --                              & 5.21                                    & --                                    & 22.86                                                  & --                                                   \\
 OQQ J0143+0151      &                    25.844 &                      1.859 &              0.334 & 44.72 $\pm$ 0.02                           & 43.54 $\pm$ 0.05                             & 1.42 $\pm$ 0.13      & $<$6.08                           & $<$6.61                         & 7.16                                    & 5.32                                  & 22.99                                                  & 23.01                                                \\
 OQQ J0149+3232      &                    27.464 &                     32.547 &              0.543 & 45.15 $\pm$ 0.02                           & 44.10 $\pm$ 0.09                             & 1.53 $\pm$ 0.97      & $<$6.78                           & $<$7.80                         & 6.53                                    & 3.63                                  & 22.95                                                  & 22.84                                                \\
 OQQ J0151+2540      &                    27.929 &                     25.671 &              0.661 & 45.07 $\pm$ 0.07                           & 44.13 $\pm$ 0.17                             & 1.25 $\pm$ 0.13      & $<$5.95                           & --                              & 8.39                                    & --                                    & 23.06                                                  & --                                                   \\
 OQQ J0231-0351      &                    37.944 &                     -3.859 &              0.450 & 44.51 $\pm$ 0.05                           & 43.74 $\pm$ 0.09                             & 1.41 $\pm$ 0.26      & --                                & $<$7.18                         & --                                      & 3.28                                  & --                                                     & 22.80                                                \\
 OQQ J0231+0038      &                    37.990 &                      0.648 &              0.488 & 45.06 $\pm$ 0.03                           & 44.14 $\pm$ 0.03                             & 1.51 $\pm$ 0.11      & $<$6.62                           & 7.61                            & 6.69                                    & --                                    & 22.96                                                  & --                                                   \\
 OQQ J0237+0448      &                    39.323 &                      4.812 &              0.647 & 44.94 $\pm$ 0.08                           & 44.07 $\pm$ 0.26                             & 1.49 $\pm$ 0.18      & $<$6.81                           & --                              & 5.90                                    & --                                    & 22.91                                                  & --                                                   \\
 OQQ J0745+4301      &                   116.493 &                     43.021 &              0.517 & 44.74 $\pm$ 0.08                           & 43.83 $\pm$ 0.08                             & 1.82 $\pm$ 0.52      & $<$6.25                           & 7.16                            & 6.77                                    & --                                    & 22.97                                                  & --                                                   \\
 OQQ J0751+4028      &                   117.913 &                     40.470 &              0.587 & 45.12 $\pm$ 0.05                           & 44.09 $\pm$ 0.06                             & 1.71 $\pm$ 0.22      & $<$6.85                           & --                              & 6.28                                    & --                                    & 22.94                                                  & --                                                   \\
 OQQ J0853+4533      &                   133.324 &                     45.554 &              0.776 & 45.89 $\pm$ 0.03                           & 44.30 $\pm$ 0.09                             & 1.21 $\pm$ 0.16      & $<$7.66                           & --                              & 6.27                                    & --                                    & 22.94                                                  & --                                                   \\
 OQQ J0911+2949      &                   137.964 &                     29.825 &              0.446 & 44.58 $\pm$ 0.06                           & 44.36 $\pm$ 0.06                             & 1.12                 & $<$7.07                           & 7.71                            & 4.32                                    & --                                    & 22.78                                                  & --                                                   \\
 OQQ J0926+6347      &                   141.639 &                     63.795 &              0.556 & 45.41 $\pm$ 0.02                           & 44.32 $\pm$ 0.04                             & 1.26 $\pm$ 0.09      & $<$6.77                           & 7.58                            & 7.24                                    & --                                    & 23.00                                                  & --                                                   \\
 OQQ J0929+3253      &                   142.306 &                     32.896 &              0.781 & 45.05 $\pm$ 0.15                           & 44.33 $\pm$ 0.06                             & 1.58 $\pm$ 0.21      & $<$7.47                           & --                              & 4.55                                    & --                                    & 22.80                                                  & --                                                   \\
 OQQ J1015+2638      &                   153.802 &                     26.643 &              0.478 & 44.52 $\pm$ 0.10                           & 43.88 $\pm$ 0.08                             & 1.49 $\pm$ 0.25      & $<$6.74                           & 7.49                            & 4.99                                    & --                                    & 22.84                                                  & --                                                   \\
 OQQ J1024+0210      &                   156.191 &                      2.170 &              0.549 & 44.59 $\pm$ 0.13                           & 43.85 $\pm$ 0.09                             & 1.77 $\pm$ 0.45      & $<$6.91                           & $<$7.72                         & 4.75                                    & 2.17                                  & 22.82                                                  & 22.62                                                \\
 OQQ J1051+1857      &                   162.751 &                     18.963 &              0.617 & 45.66 $\pm$ 0.02                           & 44.06 $\pm$ 0.09                             & 1.18 $\pm$ 0.23      & $<$7.17                           & --                              & 6.90                                    & --                                    & 22.98                                                  & --                                                   \\
 OQQ J1051+3241      &                   162.794 &                     32.699 &              0.932 & 45.94 $\pm$ 0.03                           & 44.54 $\pm$ 0.17                             & 1.11 $\pm$ 0.08      & $<$7.77                           & --                              & 6.12                                    & --                                    & 22.93                                                  & --                                                   \\
 OQQ J1116+4938      &                   169.074 &                     49.637 &              0.561 & 44.83 $\pm$ 0.06                           & 44.12 $\pm$ 0.14                             & 1.85 $\pm$ 0.42      & $<$6.95                           & $<$7.60                         & 5.28                                    & 3.19                                  & 22.86                                                  & 22.79                                                \\
 OQQ J1130+1353      &                   172.614 &                     13.894 &              0.635 & 44.93 $\pm$ 0.07                           & 44.41 $\pm$ 0.05                             & 1.04 $\pm$ 0.06      & $<$6.60                           & --                              & 6.41                                    & --                                    & 22.95                                                  & --                                                   \\
 OQQ J1156+3913      &                   179.031 &                     39.232 &              0.501 & 44.54 $\pm$ 0.09                           & 43.81 $\pm$ 0.10                             & 1.56 $\pm$ 0.39      & $<$6.09                           & $<$7.09                         & 6.66                                    & 3.60                                  & 22.96                                                  & 22.84                                                \\
 OQQ J1208+1159      &                   182.155 &                     11.994 &              0.369 & 44.77 $\pm$ 0.03                           & 44.54 $\pm$ 0.03                             & 1.08 $\pm$ 0.04      & $<$6.98                           & $<$7.06                         & 5.03                                    & 4.35                                  & 22.84                                                  & 22.92                                                \\
 OQQ J1242+5124      &                   190.621 &                     51.400 &              0.524 & 44.59 $\pm$ 0.10                           & 43.99 $\pm$ 0.07                             & 1.39 $\pm$ 0.19      & $<$6.86                           & 7.26                            & 4.86                                    & --                                    & 22.83                                                  & --                                                   \\
 OQQ J1306+4028      &                   196.613 &                     40.472 &              0.582 & 45.38 $\pm$ 0.02                           & 44.07 $\pm$ 0.13                             & 1.44 $\pm$ 0.27      & $<$7.18                           & --                              & 6.14                                    & --                                    & 22.93                                                  & --                                                   \\
 OQQ J1320+5816      &                   200.131 &                     58.278 &              0.714 & 44.99 $\pm$ 0.08                           & 44.17 $\pm$ 0.09                             & 1.46 $\pm$ 0.18      & $<$6.90                           & --                              & 5.79                                    & --                                    & 22.90                                                  & --                                                   \\
 OQQ J1346+4639      &                   206.582 &                     46.654 &              0.522 & 44.59 $\pm$ 0.06                           & 43.78 $\pm$ 0.15                             & 1.73 $\pm$ 0.60      & $<$7.01                           & 7.58                            & 4.50                                    & --                                    & 22.79                                                  & --                                                   \\
 OQQ J1412+1750      &                   213.181 &                     17.849 &              0.444 & 44.64 $\pm$ 0.04                           & 43.88 $\pm$ 0.08                             & 1.85                 & $<$4.42                           & --                              & 11.09                                   & --                                    & 23.18                                                  & --                                                   \\
 OQQ J1417+1247      &                   214.286 &                     12.794 &              0.608 & 44.88 $\pm$ 0.07                           & 44.15 $\pm$ 0.11                             & 1.30 $\pm$ 0.18      & $<$7.02                           & --                              & 5.24                                    & --                                    & 22.86                                                  & --                                                   \\
 OQQ J1443+3955      &                   220.869 &                     39.929 &              0.817 & 44.90 $\pm$ 0.16                           & 44.30 $\pm$ 0.09                             & 1.11 $\pm$ 0.11      & $<$7.20                           & --                              & 4.83                                    & --                                    & 22.82                                                  & --                                                   \\
 OQQ J1454+1440      &                   223.732 &                     14.673 &              0.577 & 45.63 $\pm$ 0.02                           & 44.15 $\pm$ 0.05                             & 1.31 $\pm$ 0.14      & $<$6.93                           & --                              & 7.42                                    & --                                    & 23.01                                                  & --                                                   \\
 OQQ J1507+5932      &                   226.885 &                     59.550 &              0.620 & 44.90 $\pm$ 0.06                           & 43.99 $\pm$ 0.09                             & 1.34 $\pm$ 0.52      & $<$6.88                           & --                              & 5.65                                    & --                                    & 22.89                                                  & --                                                   \\
 OQQ J1526+5603      &                   231.723 &                     56.066 &              0.495 & 44.52 $\pm$ 0.06                           & 43.98 $\pm$ 0.09                             & 1.42 $\pm$ 0.22      & $<$5.01                           & --                              & 9.30                                    & --                                    & 23.11                                                  & --                                                   \\
 OQQ J1538+2911      &                   234.691 &                     29.193 &              0.477 & 44.58 $\pm$ 0.05                           & 43.84 $\pm$ 0.12                             & 1.66 $\pm$ 0.37      & --                                & $<$7.03                         & --                                      & 3.86                                  & --                                                     & 22.87                                                \\
 OQQ J1540+4640      &                   235.193 &                     46.667 &              0.573 & 44.98 $\pm$ 0.03                           & 43.99 $\pm$ 0.09                             & 1.33 $\pm$ 0.22      & $<$6.97                           & $<$8.16                         & 5.61                                    & 2.22                                  & 22.89                                                  & 22.63                                                \\
 OQQ J1611+2247      &                   242.864 &                     22.787 &              0.809 & 44.98 $\pm$ 0.20                           & 44.38 $\pm$ 0.11                             & 1.18 $\pm$ 0.14      & $<$7.33                           & --                              & 4.71                                    & --                                    & 22.81                                                  & --                                                   \\
 OQQ J1611+2115      &                   242.985 &                     21.266 &              0.654 & 45.16 $\pm$ 0.05                           & 44.07 $\pm$ 0.97                             & 1.19 $\pm$ 0.10      & $<$7.32                           & --                              & 5.22                                    & --                                    & 22.86                                                  & --                                                   \\
 OQQ J1626+5049      &                   246.532 &                     50.817 &              0.522 & 44.57 $\pm$ 0.05                           & 43.48 $\pm$ 0.25                             & 1.35 $\pm$ 0.37      & $<$5.77                           & $<$6.87                         & 7.55                                    & 4.24                                  & 23.02                                                  & 22.91                                                \\
 OQQ J1629+4303      &                   247.271 &                     43.058 &              0.644 & 44.99 $\pm$ 0.07                           & 44.31 $\pm$ 0.09                             & 1.22 $\pm$ 0.08      & --                                & --                              & --                                      & --                                    & --                                                     & --                                                   \\
 OQQ J2209+3044      &                   332.427 &                     30.734 &              0.482 & 44.68 $\pm$ 0.06                           & 43.87 $\pm$ 0.06                             & 1.58 $\pm$ 0.26      & $<$6.51                           & 7.46                            & 5.99                                    & --                                    & 22.92                                                  & --                                                   \\
 OQQ J2229+2351      &                   337.440 &                     23.853 &              0.427 & 44.64 $\pm$ 0.04                           & 44.00 $\pm$ 0.10                             & 1.57                 & $<$6.49                           & 7.72                            & 5.93                                    & --                                    & 22.91                                                  & --                                                   \\
\hline
\end{tabular}
	}
\end{table}
\end{landscape}

These objects provide a valuable counterpoint to our selection, which is also based on high luminosity, but with no detectable \oiii$\lambda$5007 flux. We combine the two tables, removing any duplicates. A small minority of sources were selected by \citet{yuan_spectroscopic_2016} as having the wrong redshift fit by \sdss, and their corrected redshift is used in the processing. These are flagged in the table. The same procedure as used for the OQQs is used to match these QSO2s with \wise\ and the same luminosity cuts as outlined in Section \ref{sec:sampleselection} are applied to the combined catalog to produce the final comparison set. The sample includes only objects at \textit{z}<1.06, the range where \oiii\ can be detected in \boss.

\begin{table}
	\centering
	\caption{Process of cutting down objects. A full description of the steps is available in Section \ref{sec:sampleselection}.}
	\label{tab:selection}
	\resizebox{\columnwidth}{!}{
\begin{tabular}{llll}
\hline
         & Criteria                   & Number      & Percentage   \\
\hline
 Step 0  & Full xmatch table          & 495,003,196 &              \\
 Step 1  & $W1-W2$ $\geq$ 0.8         & 391,049     & 100.0 \%     \\
 Step 2  & xmatch $<$ 1 arcsec        & 356,665     & 91.21 \%     \\
 Step 3  & Redshift error $<$ 1\%     & 352,522     & 90.15 \%     \\
 Step 4  & No redshift warning        & 335,256     & 85.73 \%     \\
 Step 5  & $z$ in range               & 89,425      & 22.87 \%     \\
 Step 6  & Reliable WISE data         & 50,207      & 12.84 \%     \\
 Step 7  & $L_{12}$ $\geq$ $3\times10^{44}$ erg s$^{-1}$       & 36,642      & 9.37 \%      \\
 Step 8  & Class = GALAXY             & 2,033       & 0.520 \%     \\
 Step 9  & Subclass = null            & 1,175       & 0.300 \%     \\
 Step 10 & Primary science spectra    & 1,025       & 0.262 \%     \\
 Step 11 & No \oiii\ by fitting check & 125         & 0.032 \%     \\
 Step 12 & Estimated SNR $\geq$ 2     & 86          & 0.022 \%     \\
 Step 13 & Visual check               & 47          & 0.012 \%     \\
\hline
\end{tabular}}
\end{table}

\section{Results}
\label{sec:results}

\subsection{General sample properties}
\label{sec:generalresults}

\begin{figure}
	\includegraphics[width=\columnwidth]{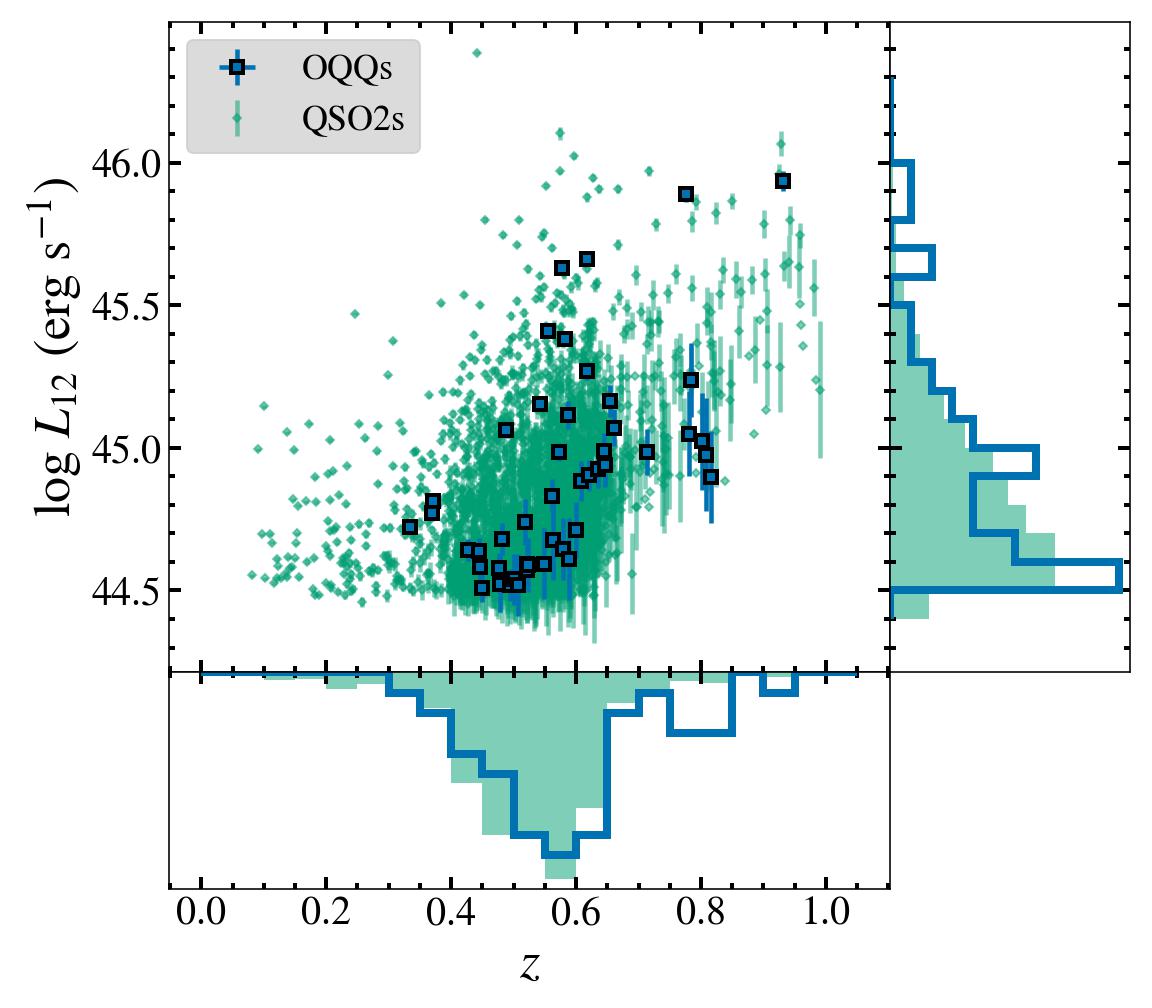}
    \caption{The \SI{12}{\micro\metre} luminosity and redshift distribution of the OQQ sample and the QSO2s.}
    \label{fig:zL12dist}
\end{figure}

The number of sources that cumulatively pass all selections is 47. We refer to this sample as `Optically Quiescent Quasars' or OQQs\footnote{With the use of \lq quasar\rq\ in this paper we are referring to {\it high luminosity} AGN rather than Type 1 AGN.}. Information on the final list is in Table \ref{tab:allobjdetails}. A sample of their spectra are in Figure \ref{fig:sdss_spectra}. The sources span a range in redshift of $0.33 \leq z \leq 0.94$ (see Figure \ref{fig:zL12dist}). No sources show significant \oiii$\lambda$5007 lines, by selection, but other low level emission lines are present in several cases.

Figure \ref{fig:zL12dist} (main panel) shows the distribution across \SI{12}{\micro\metre} luminosity and redshift for the OQQ sample and the QSO2 comparison set. Figure \ref{fig:zL12dist} (bottom) shows the redshift distribution of both the OQQ sample and the QSO2 comparison sample, and Figure \ref{fig:zL12dist} (right) shows the distribution with \lmir. The K-S statistic comparing the OQQ and QSO2 distributions is 0.073 for \lmir\ and 0.209 for redshift, with p-values 0.874 for \lmir\, and 0.007 for redshift. These results indicate that the samples are likely to come from populations with the same distribution in terms of luminosity, but with a different distribution in redshift. This is partly due to the combined selection of QSO2s from different studies, and the upgrade of the \sdss\ spectrometer - the original \sdss\ spectrometer would only allow measurements of \oiii\ up to $z\sim$0.83, hence the jump in distribution at this point. This does not invalidate their use as counterpart objects, but may have implications for any detailed analysis of evolution over time.

\subsection{Continuum Properties}
\label{sec:continuumresults}

\begin{figure}
	\includegraphics[width=\columnwidth]{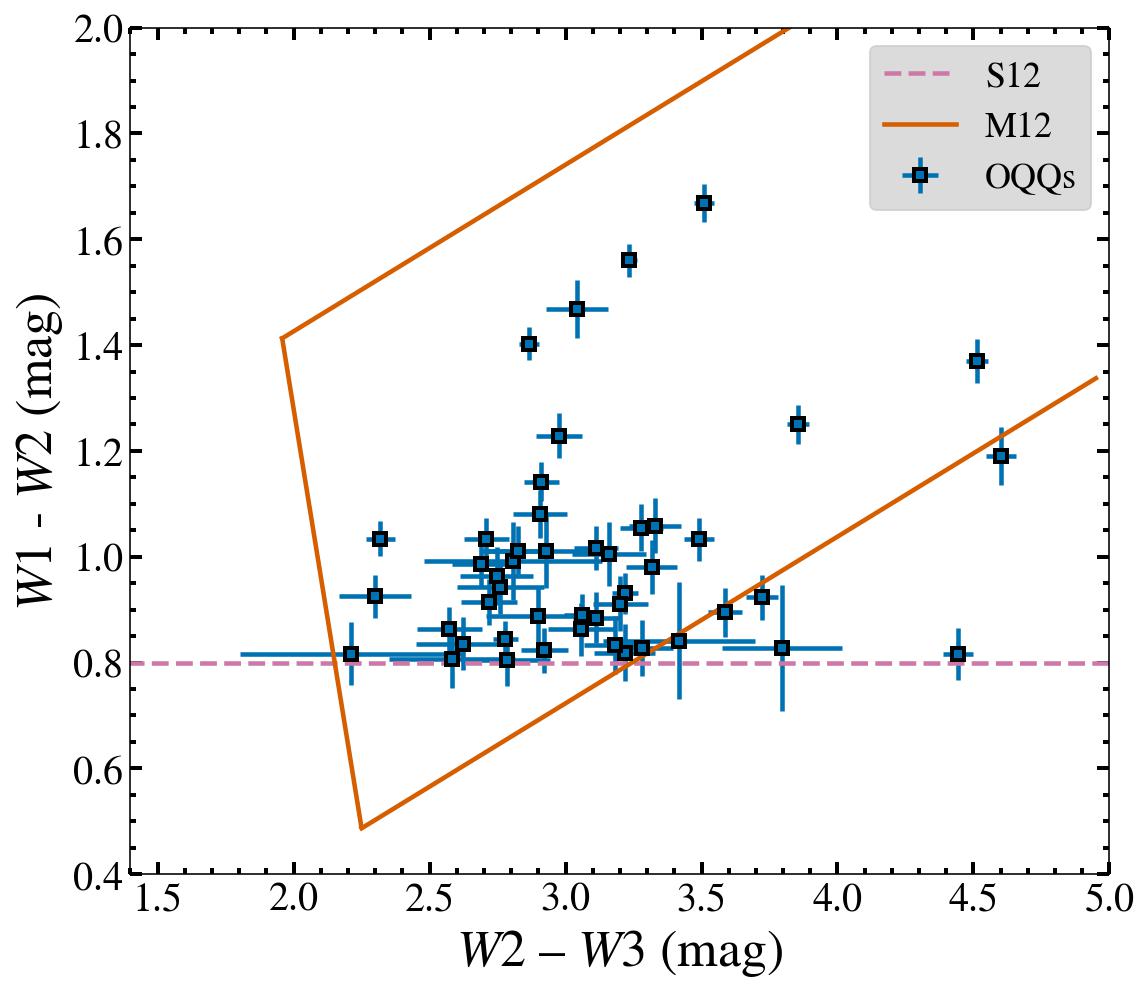}
    \caption{The \wise\ colours of the OQQ sample plotted on the standard colour-colour plane. The dashed line shows the Stern cut \citep{stern_mid-infrared_2012}, which all candidates pass by selection. The solid lines are the AGN wedge as proposed by \citet{mateos_using_2012}; most OQQ also pass according to this test.}
    \label{fig:wisecolours}
\end{figure}

With regard to the infrared properties of the sources, the \wise\ colours are plotted in Fig.\,\ref{fig:wisecolours} on the canonical \wise\ colour--colour plane. The colours of all sources are consistent with being AGN-dominated in the MIR according to the \citet{stern_mid-infrared_2012} colour cut, by selection. In addition, most of the sources also lie within the \wise\ AGN wedge proposed by \citet{mateos_using_2012}.

\begin{figure}
	\includegraphics[width=\columnwidth]{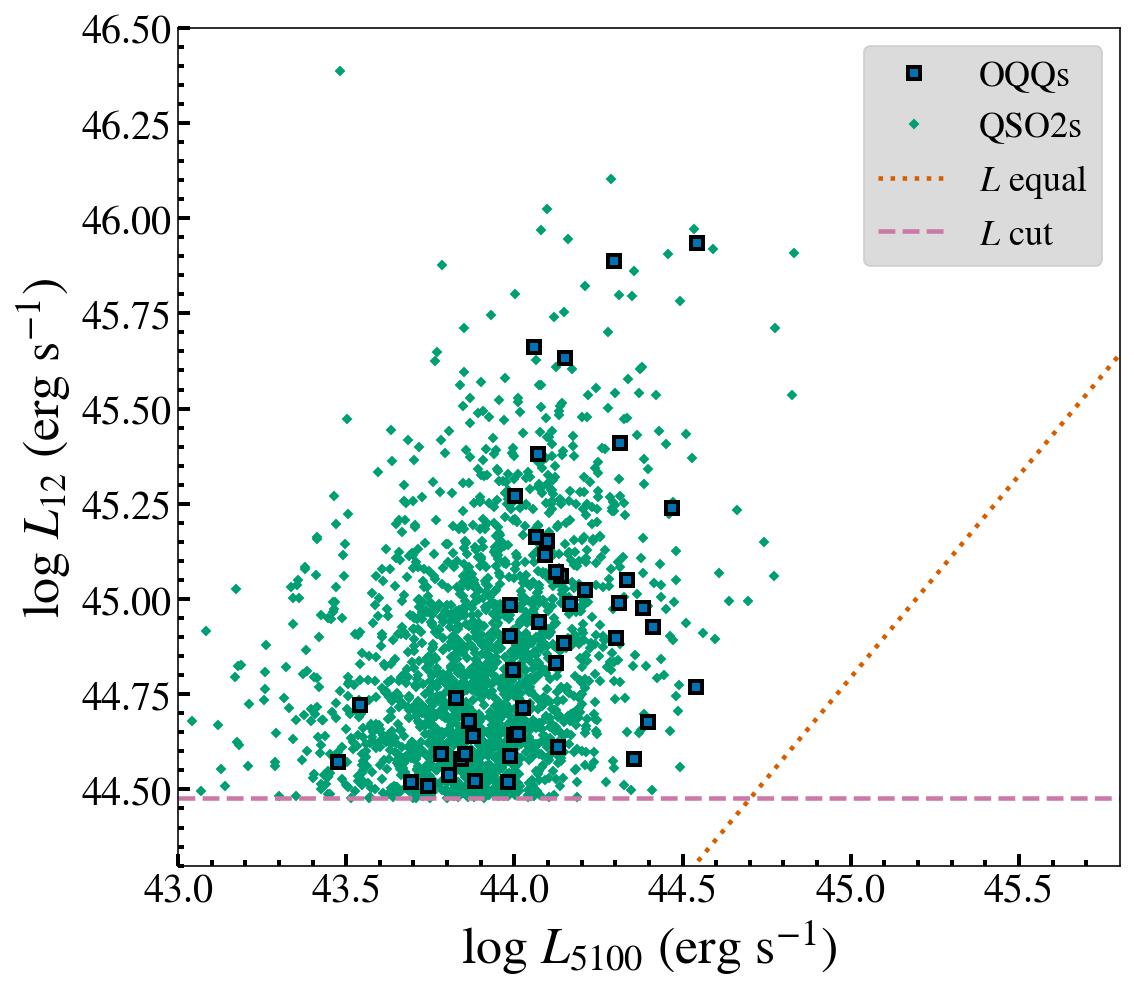}
    \caption{Comparison of \wise, \sdss\ and bolometric luminosities. OQQs are represented by outlined blue squares, and QSO2s by small green diamonds. Our luminosity threshold is shown by a horizontal dashed line. The dotted line represents where we would expect the points to lie if the IR and optical luminosities were equal.}
    \label{fig:lwise_lsdss}
\end{figure}

Fig.\,\ref{fig:lwise_lsdss} compares the \wise\ and \sdss\ luminosities of the sources. For each source, we plot the monochromatic \SI{12}{\micro\metre} and \SI{5100}{\angstrom} observed (i.e. not absorption-corrected) rest-frame luminosities (in $\lambda L_\lambda$).

The same figure also includes QSO2s: as already mentioned, QSO2s are likely to be a complementary sample to OQQs. All QSO2s with a \wise\ counterpart and MIR luminosity \lmir\ $>$\,3\,$\times$\,10$^{44}$\,erg\,s$^{-1}$ are included (as described in Section \ref{sec:qso2s}), mimicking our selection strategy, and resulting in a sample of 1,990 QSO2s. The OQQ sample is reasonably well matched to the sample of QSO2s: at 5100\,\AA\ the average luminosity of the QSO2s is slightly dimmer, and the same (to a lesser extent) at \SI{12}{\micro\metre}.

\subsection{Emission Line Properties}
\label{sec:emissionlineresults}

\begin{figure}
	\includegraphics[width=\columnwidth]{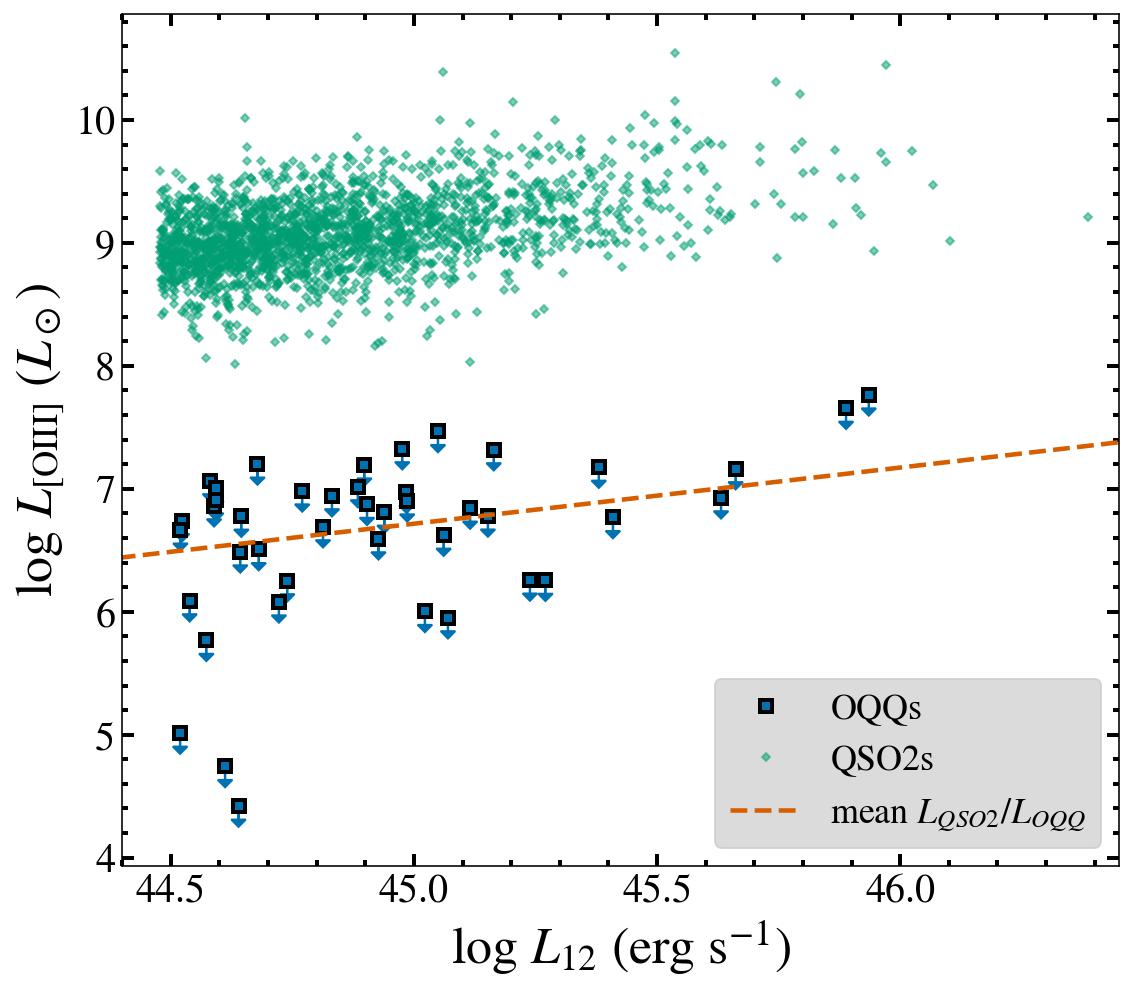}
    \caption{A comparison of the 3$\sigma$ upper limits on the \oiii$\lambda$5007 luminosity of the OQQ (large outlined blue square markers) and of the QSO2 sample (small green diamond markers), and their respective \SI{12}{\micro\metre} luminosities. The dashed line shows the slope of the relationship between QSO2 \lmir\ and measured \oiii, scaled down by the average deficit.}
    \label{fig:compareoiiiL12}
\end{figure}

Fig.\,\ref{fig:compareoiiiL12} compares the 3$\sigma$ limits on the \oiii\ line luminosity with the observed MIR power. The dashed line denotes the relation \loiii =\,4\,$\times$\,10$^{-5}$\,\lmir, which is the median ratio of the \oiii\ limits to the \SI{12}{\micro\metre} detections for the OQQs. The plot shows that our sources have a similar MIR luminosity distribution as QSO2s above the selection threshold of \lmir\,$>$\,3\,$\times$\,10$^{44}$\,erg\,s$^{-1}$. On the other hand, the \loiii\ limits of OQQs clearly lie much below the corresponding typical line detections for QSO2s.

In the following steps, we fit lines to the OQQ and QSO2 populations in order to estimate the reddening effect of the obscuring material on an average rather than individual source level. This line should not be taken as an indication of a relationship between the two properties - as we are only fitting upper limits it will be heavily effected by e.g. level of obscuration and/or intrinsic luminosity of the source. For this reason we have not made any special considerations in the fitting method for the fact that these are upper limits.

A simple linear fit to the relation between \loiii\ and \lmir\ (in log space) for QSO2s gives us the equation:
\begin{dmath}
	{\rm log}\left(\frac{L_{\rm  [O\,{\scriptscriptstyle III}]}}{L_\odot}\right)=(-37.75 \pm 0.13)+(1.04\pm 0.02){\rm log}\left(\frac{L_{12}}{\rm erg\,s^{-1}}\right)
	\label{eq:loiiil12}
\end{dmath}
This fit is based upon the ordinary least squares bisector method of \citet{isobe_linear_1990}. There is significant scatter of 0.11 dex and a Spearman rank correlation coefficient of 0.40 indicates that the relation is not very strong. Nevertheless, this serves as a guide for comparison to OQQs. 

\begin{figure}
 	\includegraphics[width=\columnwidth]{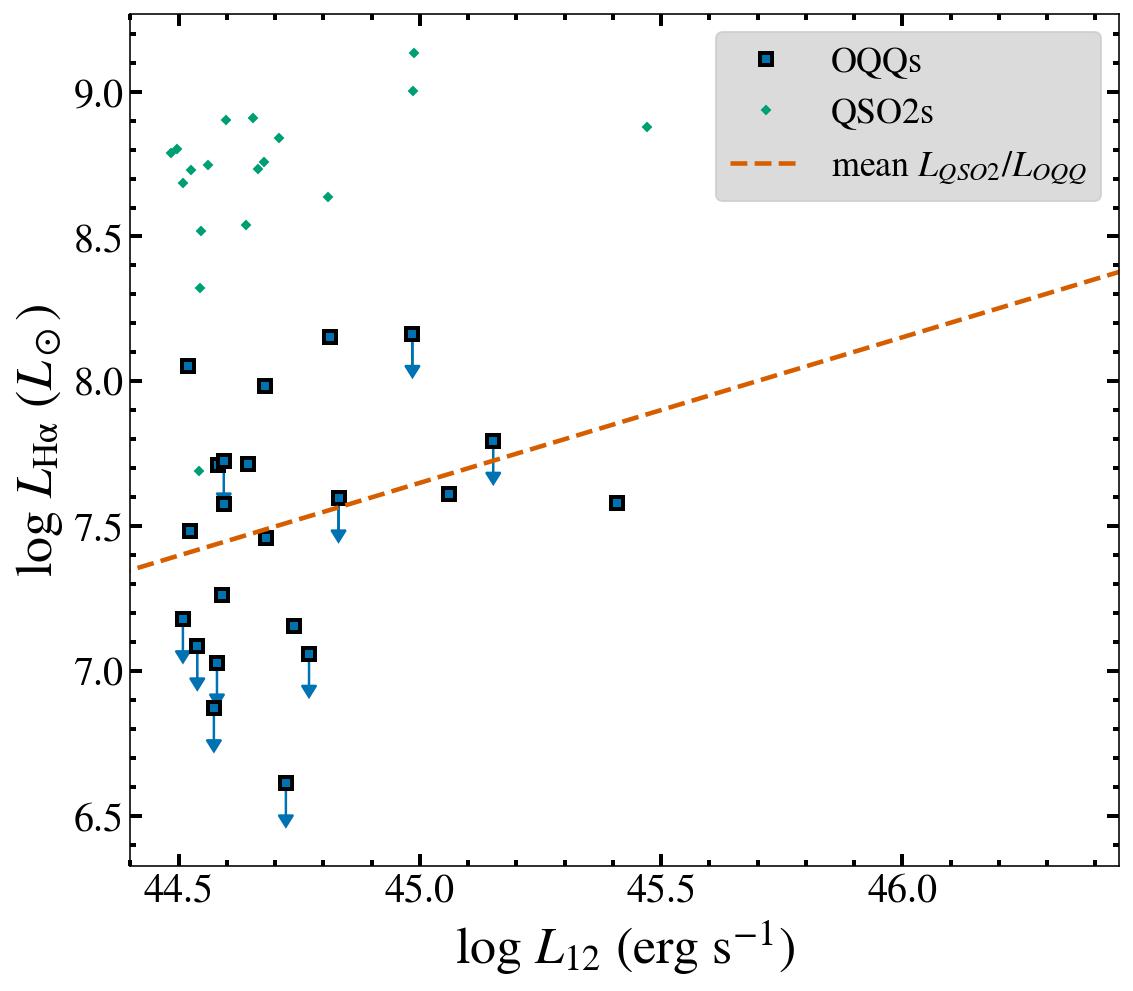}
    \caption{A comparison of the \ha$\lambda$6562 luminosity (or its upper limit - out of the sample of 62 OQQs, 17 have detected \ha\ emission lines), with OQQs (large blue square markers) and QSO2s (small green diamond markers), and their respective \SI{12}{\micro\metre} luminosities (for those objects where \ha$\lambda$6562 is not redshifted out of the \sdss\ range). The dashed line shows the slope of the relationship between QSO2 \lmir\ and measured \ha, reduced by the average deficit.}
    \label{fig:lhal12}
\end{figure}

The dotted line denotes an \oiii\ deficit of $\sim$253\,$\times$ from the above fit. This is the mean deficit of our OQQs relative to the expected intrinsic \oiii\ luminosity if they are typical quasars similar to QSO2s. Since these are upper limits, this should not be considered a true fit to the data, but merely allows us to estimate the scale of obscuration present. The actual deficits could be much larger, and potentially show a different relationship to \SI{12}{\micro\metre} luminosity. One can perform a similar comparison using the \ha\ emission line, as it is less affected by reddening. The result is shown in Fig.\,\ref{fig:lhal12}.

The corresponding relation for QSO2s is
 
\begin{dmath}
	{\rm log}\left(\frac{L_\ha}{\rm erg\,s^{-1}}\right)=(-10.44\pm 2.30)+(1.18\pm 0.29){\rm log}\left(\frac{L_{12}}{\rm erg\,s^{-1}}\right)
	\label{eq:lhal12}
\end{dmath}

QSO2 \oiii$\lambda$5007 fluxes from the source papers \citep{reyes_space_2008, yuan_spectroscopic_2016} were used, but \ha$\lambda$6562 were drawn from the {\tt emissionLinesPort} table. Since we are only using the QSO2 sample for comparison, we did not perform our own \ha\ line flux measurements. Above $z$\,$\approx$\,0.35, when redshifted \ha\ is shifted to $\gtsim$\,9000\,\AA, the sky noise can often render lines difficult for the pipeline to measure correctly. Therefore, we restricted the QSO2 sample to below this redshift for extracting robust automated \ha\ fluxes. This results in few QSO2s for the above fit, all of which have a significant \ha\ detection without having been specifically selected on this line, as one would expect for a standard quasar spectrum. The scatter in this case is 0.10 dex with a Spearman rank correlation coefficient of 0.48.

The OQQs have a median distribution (either a detection or a limit) consistent with an \ha\ deficit of $\approx$\,27. This is less than the deficit found from \oiii\ partly because of the lower reddening affecting \ha, and partly because redshifted \ha\ typically lies in the reddest portions of the spectra with strong noise from sky emission lines resulting in less sensitive non-detection flux limits. Another possibility is that \ha\ has a larger fraction of its flux coming from star formation, while \oiii\ is likely primarily from the AGN.

\subsection{Spectral Energy Distribution (SED) fitting}\label{sec:agnfitter}

\begin{figure*}
	\includegraphics[width=\textwidth]{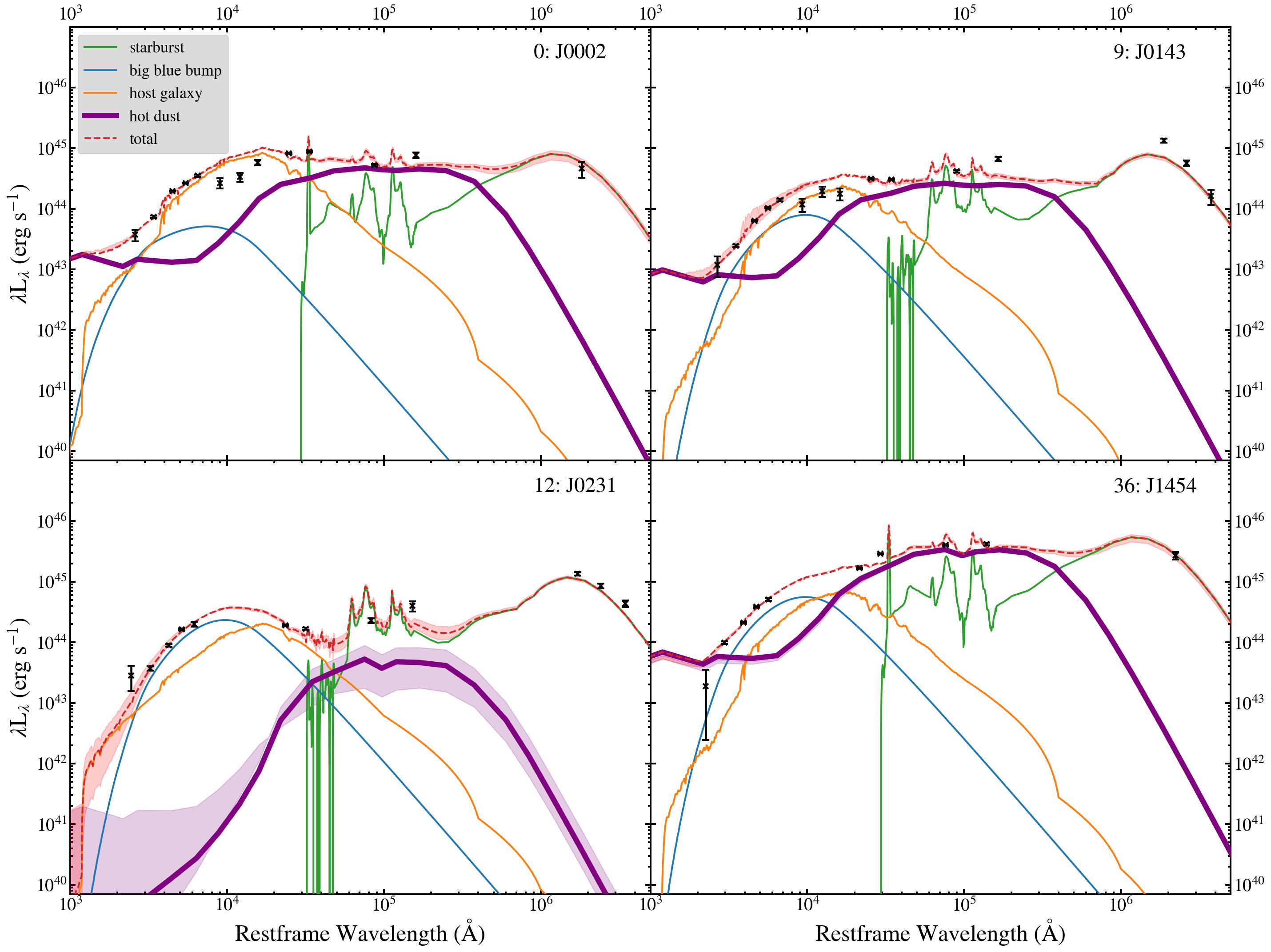}
    \caption{\agnfitter\ results for four candidate OQQs with long wavelength IR data. The hot dust component - that most indicative of an AGN - is highlighted (thick purple line). Errors are shown for this component and the total only. Photometric bands used in this plot (in addition to \sdss\ \textit{ugriz} and \wise\ \textit{W1-4}) were as follows: OQQ~J0002-0025 all \twomass\ and \SI{250}{\micro\metre}; OQQ~J0143+0151 all \twomass\ and the three {\em Herschel} bands stated in footnote~\ref{footnote:herschelbands}; OQQ~J0231-0351 three {\em Herschel} bands; OQQ~J1454+1440 \SI{350}{\micro\metre} only.}
    \label{fig:agnfitter}
\end{figure*}

We use \agnfitter\ \citep{calistro_rivera_agnfitter_2016} to fit the SEDs of a selection of OQQs and determine contributions to the observed emission from different components of the AGN/galaxy system. This method uses the multiwavelength data available to find the most likely weighted contributions to the overall SED, using Markov Chain Monte Carlo (MCMC) to fit from a large library of models. The most important parameter for our analysis is the contribution to the SED from the AGN obscuring material. We select the candidates with available long wavelength data (\twomass\footnote{\SI{1.235}{\micro\metre}, \SI{1.662}{\micro\metre}, \SI{2.159}{\micro\metre}.}, {\em Herschel}, in addition to all four \wise\ bands, present in all OQQs by selection), as this is key to constraining some components; this totals only 9 sources\footnote{\label{footnote:herschelbands}{\em Herschel} bands used were any available out of \SI{250}{\micro\metre}, \SI{350}{\micro\metre}, and/or \SI{500}{\micro\metre}. Of these nine sources: two have \twomass\ only; five have some combination of \twomass\ and {\em Herschel}; two have {\em Herschel} only.}. Figure~\ref{fig:agnfitter} presents the results from four OQQs that have reasonably well-constrained results. After experimenting with different MCMC lengths, we selected one sufficient to achieve an auto-correlation time indicating convergence for the OQQ AGN-heated dust component (see \citealp{foreman-mackey_emcee_2013} and \citealp{calistro_rivera_agnfitter_2016} for details). Settings used were: 100 walkers; two burn in sets (10,000 steps each); set length 200,000.

In some cases the hot dust component (e.g. the top two plots in Figure~\ref{fig:agnfitter}) rises towards the blue end. This would not occur in a purely physical model component, but it is an empirical template based on combinations of AGN SEDs from \citet{silva_connecting_2004}, and the range of values these can take can be seen in Figure 1 of \citet{calistro_rivera_agnfitter_2016} and Figure 1 of \citet{silva_connecting_2004}; they are based on \nh\ and assumed AGN type. As can be seen in these figures, the peak remains unchanged, varying only with normalisation. With further data into the blue and near-UV, we may be able to take this as some indication that the OQQ in question is lightly obscured, or at least has some accretion disc emission visible, but currently degeneracy with \lq host galaxy\rq\ emission (solid orange line) makes this conclusion premature. Nevertheless, while the general results should be used with caution, the key outcome is that a high normalisation of the \lq hot dust\rq\ component (see Figure \ref{fig:agnfitter}, thick purple line) is required to reproduce the MIR part of the OQQ SEDs regardless of the tail shape (i.e. the peak of the hot dust component at $\sim$\SI{6}{}--\SI{20}{\micro\metre} in the rest frame). We conclude it is very likely each of these objects contains an AGN bright enough to heat its obscuring dust to produce this component. \agnfitter\ makes two separate estimates of star formation rate (SFR) - one in the optical, and one in the FIR. The FIR SFR provides an estimate of potential obscured star formation, but we note that very few candidates have reliable data in that region, so we cannot make any conclusions about the whole OQQ population. This estimate (and that of stellar mass) is obtained from host galaxy template parameters (for more detail, see \citealt{calistro_rivera_agnfitter_2016}).

\subsection{4000 \AA\ break} \label{sec:d4000}

\begin{figure}
	\begin{center}
	\includegraphics[width=\columnwidth]{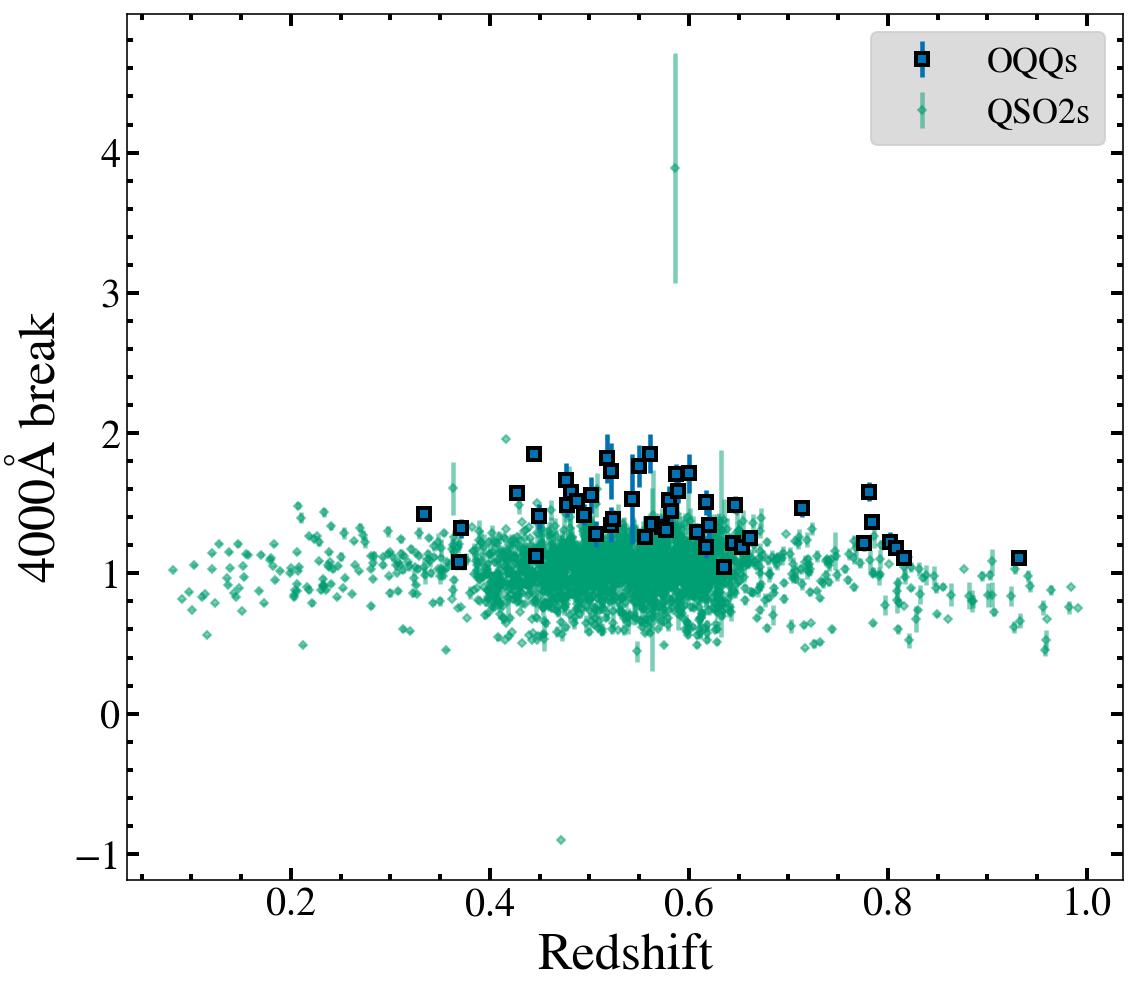}
	\end{center}
    \caption{\SI{4000}{\angstrom} break data. Blue squares are the OQQs, with error bars shown for those with D4000 significantly greater than 1. Green markers are the QSO2 comparison sample. The average D4000 is clearly higher for OQQs than QSOs, implying a difference in the host galaxy population. However, this may be a selection rather than intrinsic effect.}
    \label{fig:breakdata}
\end{figure}

The 4000\,\AA\ break (D4000 from this point on), an indicator of a significant population of old stars, was calculated as described in \cite{balogh_differential_1999}, as the ratio between red (4000\,\AA\ - 4100\,\AA) and blue (3850\,\AA\ - 3950\,\AA) continuum fluxes. These ranges are slightly more restrictive than in previous works \citep[e.g.][]{hamilton_spectral_1985}, as this was found to improve on the repeatability of D4000 between separate measurements, and shows less sensitivity to reddening. A large 4000\,\AA\ break could be indicative of significant host galaxy light dilution.  The values are found in Table \ref{tab:allobjdetails} and shown in Figure~\ref{fig:breakdata}: 44 out of 47 (94\%) objects have a significant D4000. The median value is 1.41 (including only the significant breaks, defined as >3 standard deviations above unity). The QSO2 sample found only 811/1988 (41\%) had a significant D4000.

\section{X-ray observations}
\label{sec:xrayobs}

In the absence of AGN spectroscopic signatures in the optical, X-ray observations arguably provide the best evidence for the presence of an AGN independent of the MIR signatures. We checked the {\sc heasarc} archive\footnote{\url{heasarc.gsfc.nasa.gov/docs/archive.html}} for pointed or serendipitous observations by X-ray satellites, and discovered several OQQs observed during targeted observations of other nearby objects. We also obtained joint \xmm\ and \nustar\ \citep{harrison_nuclear_2013} data for our prototypical OQQ (see Section~\ref{sec:discussion}), and these results are discussed in Section \ref{sec:J0751xray}. Details of a serendipitous observation with sufficient counts for fitting are presented in Section~\ref{sec:J1051xray}.

The \xmm\ mission has one of the best combinations of effective area and angular resolution of current X-ray missions, and is therefore an ideal tool to help us analyse the X-ray properties of OQQs. \chandra, which has very low background noise, can also be valuable in detecting dim sources. We also examine \xrt\ data - this is generally less sensitive, but in cases where no \xmm\ or \chandra\ data is available \xrt\ can still be useful to place limits on the X-ray emission.  
For detected sources, we extracted source and background spectra using standard {\sc ftools} tasks and recommended reduction steps for each instrument, and analysed the data using the \xspec\ package \citep{xspec}. We then assessed the false detection probability based on counts in the source region relative to counts in a background region, and this derived either estimated fluxes or upper limits for each target \citep[see e.g.][]{lansbury_nustar_2017a}.
Figures \ref{fig:xrayL12comp} and \ref{fig:xrayOIIIcomp} show the results of analysis of data from \xmm, \chandra, and \xrt\ for OQQ candidates with serendipitous observations - in total 11 separate OQQs (See Table \ref{tab:oqq_xray}). These are discussed further in Section~\ref{sec:xraydiscussion}.

\subsection{X-ray Observations of OQQ~J0751+4028}
\label{sec:J0751xray}

We obtained targeted observations of our prototypical candidate with the aim of confirming its AGN nature and constraining its properties. Following standard procedures with \xspec\ we modelled the source, finding that it is under-luminous at 2--10~keV (compared to the IR-predicted intrinsic luminosity), and appears to be lightly obscured (\nh$\sim$10$^{22}$ cm$^{-2}$). This work is summarised here: for details of the observations and procedures, see \citet{greenwell_xmm_2022}.

\begin{figure}
	\includegraphics[width=\columnwidth]{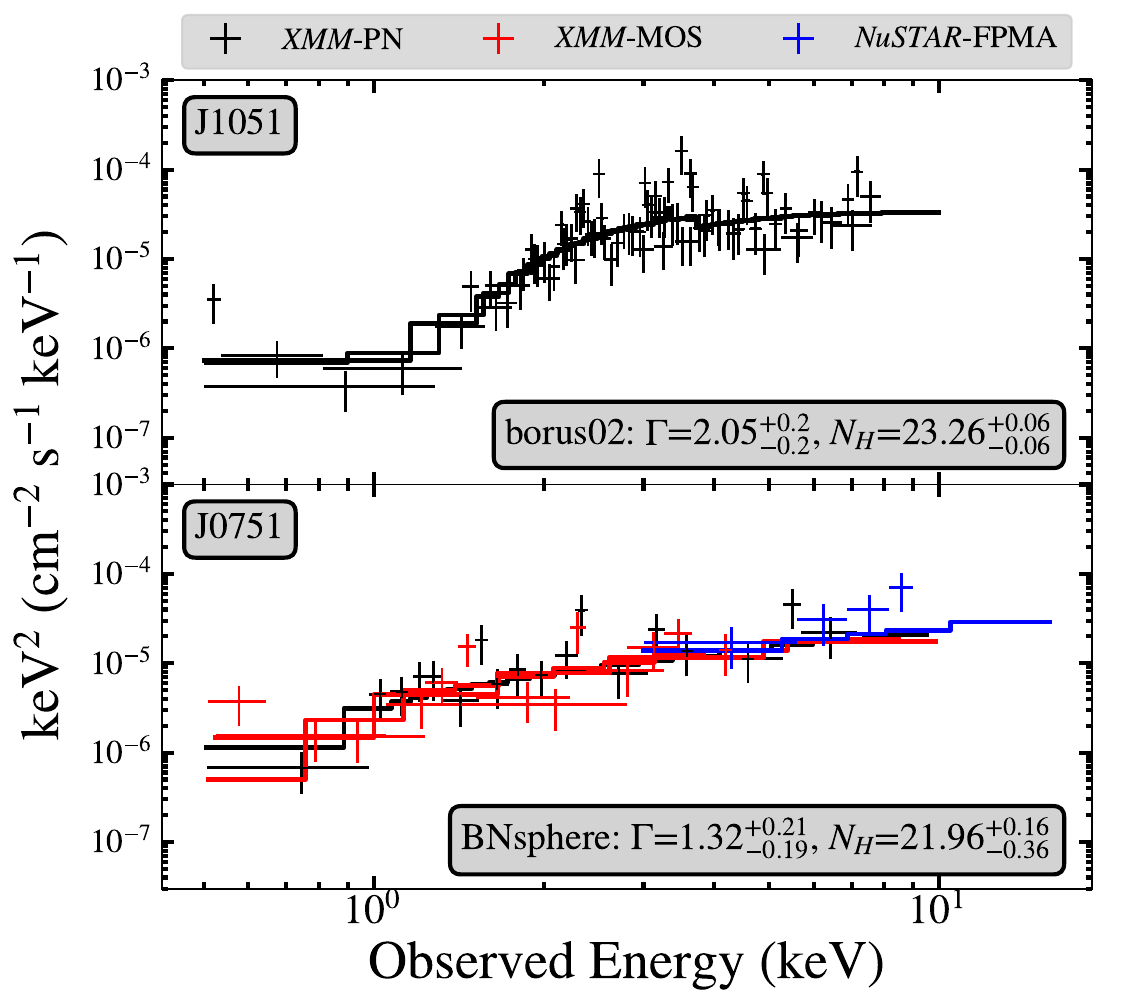}
    \caption{X-ray spectra of two OQQs. Top panel: OQQ~J1051+3241, fitted with {\sc borus02}; bottom panel OQQ~J0751+4028, fitted with {\sc BNsphere}. The sources have several differing properties (e.g. \nh, intrinsic spectral shape) but have underluminous intrinsic luminosities (see Figure~\ref{fig:xrayL12comp}).}
    \label{fig:xrayobscomp}
\end{figure}

We fit several models, starting with an absorbed power law ({\sc cabs*zwabs*pow}), moving on to more physically realistic models ({\sc mytorus}, \citealt{murphy_x-ray_2009}; {\sc BNsphere}, \citealt{brightman_xmm-newton_2011}) and finally a more complex model involving a lightly obscured leak in an otherwise Compton thick sphere of obscuration. Using Bayesian X-ray Analysis \citep[BXA; ][]{buchner_x-ray_2014} we can compare the results and determine the model most likely to have produced the data. We find that {\sc BNsphere} is favoured - a spherical obscuration model that implies a \lq cocooned\rq\ obscuring structure may be a plausible explanation for the X-ray appearance of OQQ~J0751+4028. This result is shown in Figure~\ref{fig:xrayobscomp} (bottom panel).

\begin{figure}
	\includegraphics[width=\columnwidth]{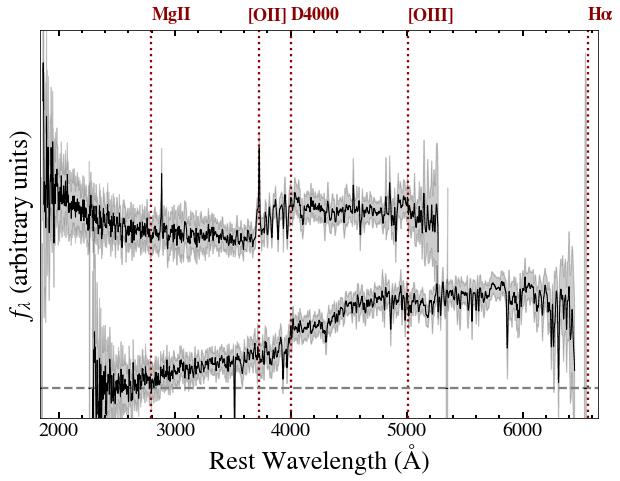}
    \caption{\sdss\ spectra of OQQ~J1051+3241 and OQQ~J0751+4028. J1051 shows a slight blue rise which may indicate that there is either some view of the accretion disc or recent star formation (see Appendix~\ref{app:vischeck}), whereas J0751 shows the continuum shape and lack of emission lines of typical OQQs.}
    \label{fig:xrayobscomp_optspec}
\end{figure}

The \sdss\ spectrum of OQQ~J0751+4028 is shown in Figure \ref{fig:xrayobscomp_optspec} (bottom spectrum), and (by selection) shows typical properties for an OQQ (compared to examples from the general sample in Figure~\ref{fig:sdss_spectra} and online). For more detail see \citet{greenwell_candidate_2021}.

Figure \ref{fig:xrayL12comp} shows OQQ~J0751+4028 in context with the serendipitous observations and the empirical relationship between \SI{12}{\micro\metre} and intrinsic 2-10~keV luminosity. The unobscured rest frame 2-10~keV luminosity is 4.39 $\times$ 10$^{43}$~erg~s$^{-1}$, approximately a factor of 6 lower than expected given the \SI{12}{\micro\metre} luminosity, but sufficient to categorically demonstrate the presence of an AGN. The photon index is harder than usual for an AGN ($\Gamma$=1.32$^{+0.21}_{-0.19}$), and obscuration is approximately Compton-thin but present (log (\nh\ / cm$^{-2}$) = 21.96$^{+0.16}_{-0.36}$). If $\Gamma$ is fixed to a more usual AGN-like value of 1.9, the obscuration increased only slightly to log ($N_{\rm H}$ / cm$^{-2}$) = 22.3.

\subsection{Serendipitous X-ray observations of OQQ~J1051+3241}
\label{sec:J1051xray}

One serendipitous OQQ had sufficient \xmm\ counts across two observations (0781410101 and 0781410201; both only observed the source with the PN detector) to be analysed further with \xspec. We fit this source (OQQ~J1051+3241) with several models, as was done with the targeted observations of OQQ~J0751+4028 \citep[see Section~\ref{sec:J0751xray} and][]{greenwell_xmm_2022}. We find the best fit model to be {\sc borus02} \citep{balokovic_new_2018}, at fixed angle of inclination assuming viewing through the torus ($\cos{\theta}=0.05$). We do not find a significant Fe~K$\alpha$ emission line, and allowing the angle of inclination to vary does not favour any angle strongly. This object is more obscured and has a higher $\Gamma$ than J0751: log (\nh\ / cm$^{-2}$) = 23.26$\pm0.06$, $\Gamma$=2.05$\pm0.2$ (see Figure~\ref{fig:xrayobscomp}, top panel). For all sources (with the exception of our targeted observation; Section \ref{sec:J0751xray}), we have assumed the same photon index as for this source, and used WebPIMMS\footnote{\url{https://heasarc.gsfc.nasa.gov/cgi-bin/Tools/w3pimms/w3pimms.pl}} to estimate the luminosity of each source, given the net counts (or upper limit) observed. We find that the detected OQQs are underluminous compared with expectations. If the non-detected OQQs are equally underluminous, they may still be detectable with sufficiently long observations.

The \sdss\ spectrum of OQQ~J1051+3241 is shown in Figure \ref{fig:xrayobscomp_optspec} (top spectrum). In this case it shows a slight rise towards the blue end, implying that this is not as strictly \lq typical\rq\ of an OQQ as OQQ~J0751+4028.

\subsection{Discussion of X-ray results}\label{sec:xraydiscussion}

\begin{figure}
	\includegraphics[width=\columnwidth]{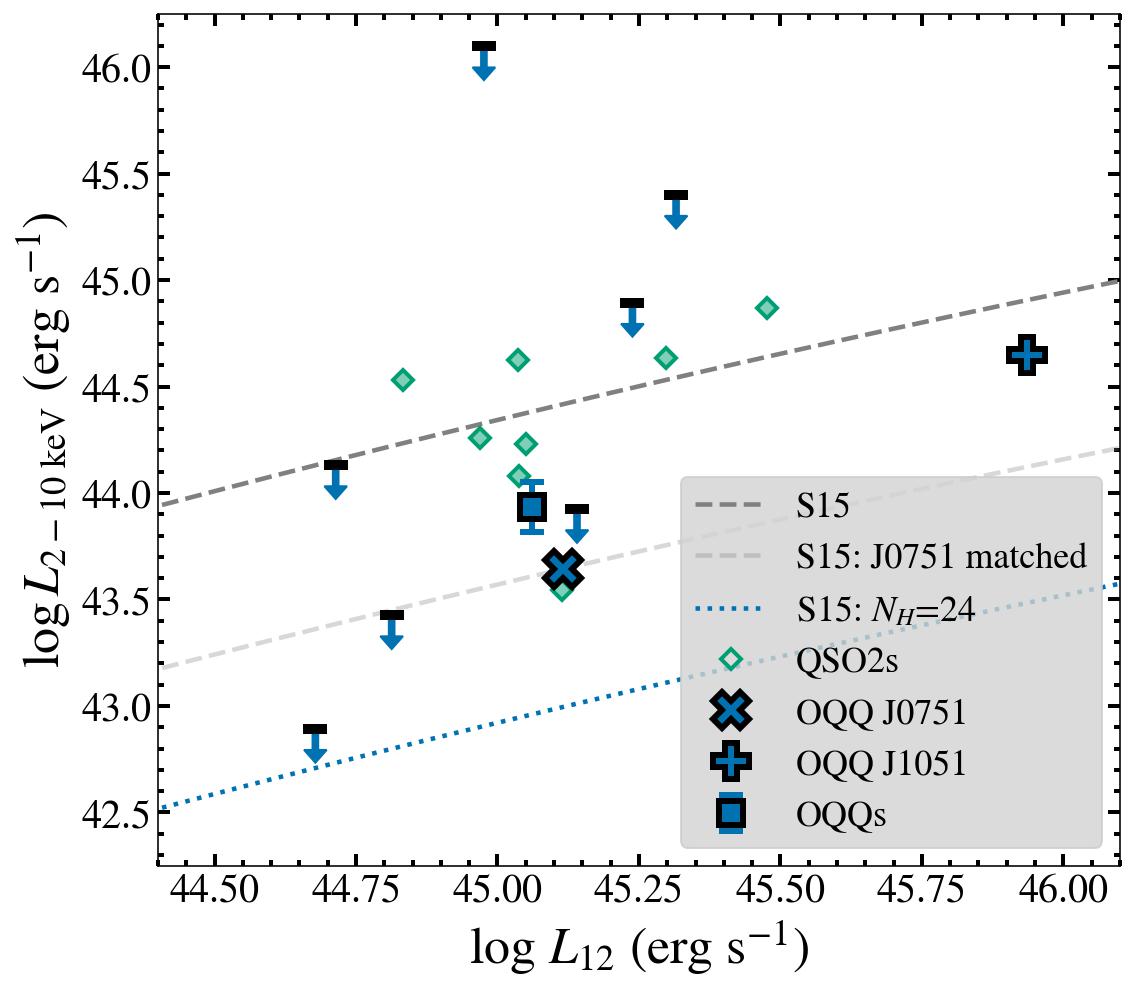}
    \caption{Observed 2--10\,keV luminosities of the OQQs (blue markers), plotted against their \SI{12}{\micro\metre} luminosities. OQQ~J0751+4028 is shown as a large cross marker. For comparison, available QSO2 luminosities are shown as diamonds (see Table \ref{tab:qso2_xray}). We also show MIR predicted values for intrinsic luminosities \citep[dashed line;][]{stern_x-ray_2015}. If this relationship is scaled to the luminosity of J0751 \citep[pale grey; ][]{greenwell_xmm_2022}, the other detected OQQs are more consistent with this expectation. Upper limits may be better explained by a Compton thick scenario: dotted blue line shows the obscuration of the S15 relation by Compton thick material.}
    \label{fig:xrayL12comp}
\end{figure}

\begin{figure}
	\includegraphics[width=\columnwidth]{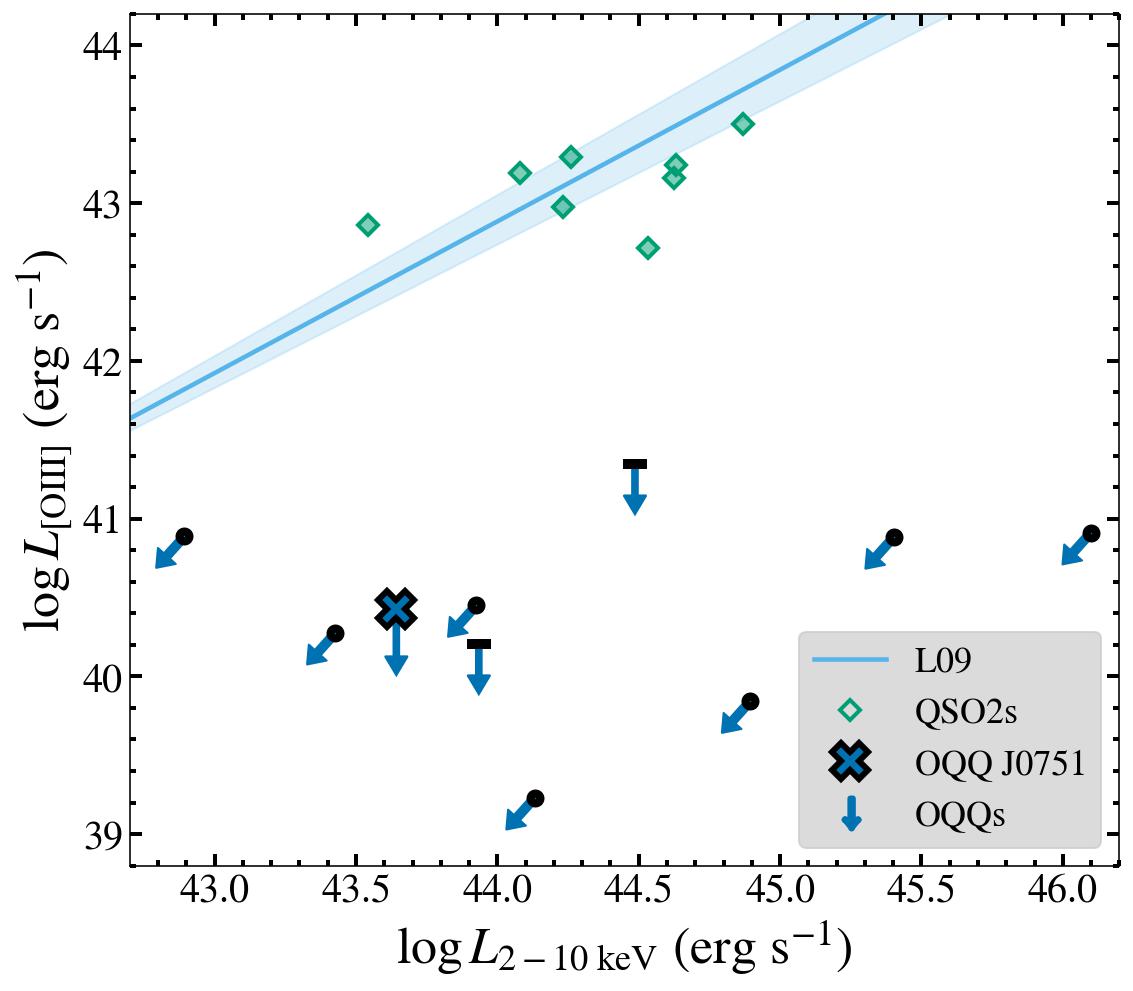}
    \caption{2--10\,keV luminosities of the OQQs (blue markers), plotted against their \oiii\ upper limit luminosities.  OQQ~J0751+4028 is shown as a large cross marker. For comparison, available QSO2 luminosities are shown as diamonds (see Table \ref{tab:qso2_xray}). We also show \oiii\ predicted values for intrinsic luminosities \citep[solid line;][]{lamastra_bolometric_2009}.}
    \label{fig:xrayOIIIcomp}
\end{figure}

Figure \ref{fig:xrayL12comp} shows the comparison between X-ray and \SI{12}{\micro\metre} luminosity, including the predicted relationship between intrinsic X-ray and nuclear \SI{12}{\micro\metre} luminosity from \cite{asmus_subarcsecond_2015} and between X-ray and \SI{6}{\micro\metre} luminosity \citep{stern_x-ray_2015}; for the latter, \SI{6}{\micro\metre} luminosities were converted to \SI{12}{\micro\metre} luminosities using a relationship derived from the QSO template of \citet{hao_distribution_2007}. The OQQ X-ray luminosities are slightly lower than the relations would predict, but this may be due to the difference between observed and intrinsic IR luminosities - if the OQQs are heavily obscured (whether from the circum-nuclear material or larger scale obscuration), their intrinsic luminosities will be higher. OQQ IR luminosities refer to the total galaxy emission, not the nuclear emission alone, as with current data there is no reliable way to extract or convert this for individual objects. Figure \ref{fig:xrayL12comp} shows where obscuration-reduced MIR-predicted X-ray luminosities \citep{asmus_subarcsecond_2015, stern_x-ray_2015} would fall. This would imply that most OQQs lie somewhere between Compton thick and unobscured. However, if they are intrinsically similar to OQQ~J0751+4028 \citep{greenwell_xmm_2022} the obscuration depth is low for an obscured AGN (\nh$\sim$10$^{22}$~cm$^{-2}$), and thus the intrinsic values for the OQQs may be only slightly higher. Figure~\ref{fig:xrayOIIIcomp} shows the relationship between \oiii\ and X-ray luminosity found by \citet{lamastra_bolometric_2009}, with the QSO2s falling where expected. Upper limits on \oiii\ luminosity place the OQQs significantly below expectations, reinforcing the disconnect between observed \oiii\ and intrinsic properties.

\section{Discussion}
\label{sec:discussion}

The combination of \wise\ all-sky MIR photometry and \sdss\ optical spectroscopy have enabled us to identify an interesting class of sources showing all the characteristics of luminous AGN in the MIR (based upon their colours and luminosities) but no obvious signatures of the corresponding AGN activity in the optical. 

Multiple physical scenarios can explain these observations, principally based on whether the optical AGN signatures are obscured, or intrinsically absent. In the former, we can consider the OQQ in terms of the spatial scale of its obscuration. As opposed to the putative donut shape envisaged by the zeroth order AGN unified torus scheme, full or near full sky covering by dusty \lq cocoons\rq\ could easily explain the optical absence of emission signatures by not allowing an extended NLR to form or be ionized in the first place. Such cocoons would still reprocess the intrinsic AGN emission to the MIR, and the high \wise\ luminosities (along with bright hot dust component, where SED fitting is available) of these sources strongly suggest that the underlying power sources are luminous quasars, as does the MIR comparison with \sdss-selected QSO2s. For the latter scenario, AGN emission lines are not currently produced by the OQQ. We can envisage this as a \lq young\rq\ AGN - recently switched on, with the AGN radiation not yet ionising the NLR, and consequently no narrow emission lines. It is still likely that this is an obscured AGN, i.e. that it is being viewed through a dusty torus, which would hide any broad emission lines (produced closer to the supermassive black hole (SMBH), and therefore earlier in the switching on process). As in the \lq cocooned\rq\ picture, reprocessed intrinsic emission can still produce the observed IR properties.

We previously identified a prototypical example of this class, described in \citet{greenwell_candidate_2021} and \citet{greenwell_xmm_2022}. We found that it indicated a new subclass of AGN that does not easily fit into current models:

\begin{itemize}
	\item The spectral shape of the optical continuum is that of a typical galaxy, including showing a strong 4000 \AA\ break.
	\item Strong upper limits on emission lines that may indicate the presence of an AGN - there is no \oiii\ by selection. Others may be present at a very low level.
	\item The \wise\ IR colours are firmly within the AGN wedge \citep{mateos_using_2012}, a stricter criterion than that required for the main sample selection.
	\item The \SI{12}{\micro\metre} luminosity is very high, consistent with the quasar regime: there is a high chance of a hidden AGN being present, and it is not likely that this comes from a different mechanism.
	\item X-ray luminosity is luminous enough to clearly indicate the presence of an AGN, but under-luminous compared to expectations from \SI{12}{\micro\metre} luminosity (see \ref{sec:J0751xray}).
    \item An upper limit on radio flux is available from \first\ \citep{becker_first_1995}: <0.695 mJy/beam.
\end{itemize}

\subsection{Optical Reddening}
\label{sec:reddening}

(Non)detection of emission lines can be used to place constraints on the optical depth of the theoretical cocoons, if present (see Section \ref{sec:cocoon}). We assume that the QSO2s have low extinction to their NLRs (an approximation), and that therefore the relationship between \SI{12}{\micro\metre} and \oiii\ luminosity (\ha\ luminosity) (equations \ref{eq:loiiil12}(\ref{eq:lhal12}) above) can be used to predict the approximate intrinsic flux of a cocooned source at the same \SI{12}{\micro\metre} luminosity (assuming that all OQQs had \oiii\ signals at the upper limit of detection. The extinction \av\ can be calculated using this predicted flux, the actual measured flux, and the standard Milky Way extinction law of \citet{cardelli_relationship_1989} with an $R_{\rm V}$\,=\,3.1. The resultant values of the extinction \av\ as derived from the two lines are listed in Table \ref{tab:allobjdetails}. Since QSO2s are also likely to suffer mild NLR extinction \citep{reyes_space_2008}, these should be regarded as lower limits on the true \av. The  \av\ limits measured from the \oiii\ non-detections range over 3.4--10.5\,mag, with a median value of 4.69.

What would be the corresponding column density of gas for this extinction? A factor of 100 deficit in \oiii\ does not need extreme columns. Our median extinction \av\,=\,4.69\,mag is equivalent to a neutral gas column of \nhperp\,$\sim$\,7\,$\times$\,10$^{22}$~cm$^{-2}$ along the galaxy axis, assuming the gas--to--dust ratios found in AGN environments by \citet{maiolino_dust_2001}. The parameter \nhperp\ is used to denote the fact that these gas columns are relevant for the NLR, which is likely to probe angles out of the direct l.o.s., thus making the distinction from the \nhlos\ measured from X-rays in the previous section. Further speculations on the gas columns or spatial distributions with only two detected objects would be premature. The \nhperp\ values are lower limits inferred from the limits on the optical extinction, so it is unexpected that the \nhlos\ would be lower. %Furthermore, gas is expected to be present on nuclear scales smaller than the dust sublimation radius so it is natural that the X-ray--inferred \nhlos\ values be larger than \nhperp\ values inferred from dust extinction in the optical.

\begin{figure}
	\includegraphics[width=\columnwidth]{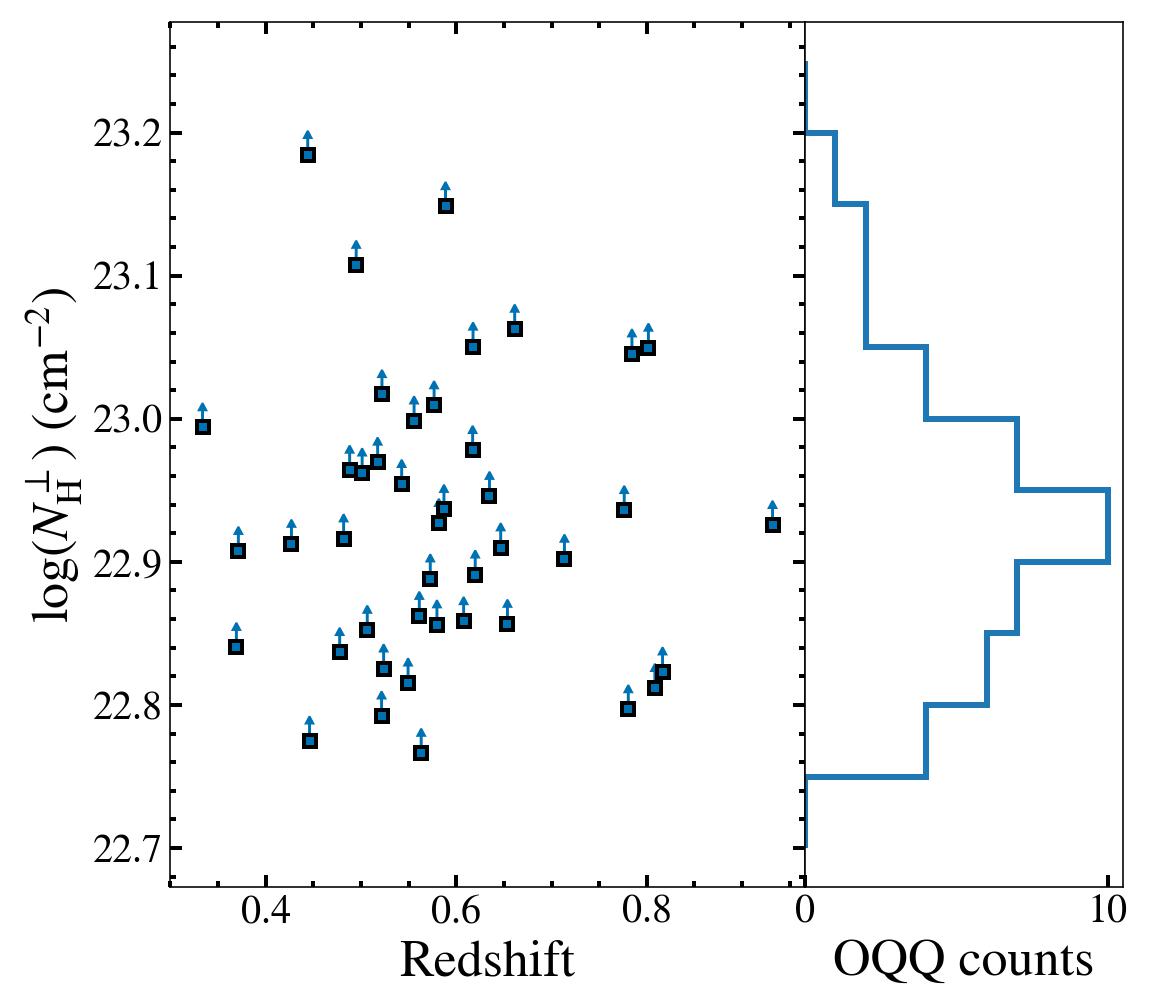}
    \caption{Inferred \nhperp\ values from \oiii\ (i.e. obscuring the NLR, distinct from the \nhlos\ related to X-ray obscuration).}
    \label{fig:inferredNH}
\end{figure}

Figure ~\ref{fig:inferredNH} shows the \nhperp\ values inferred from \oiii\ extinction. In a small number of cases, this can also be calculated using \ha\ (see Table~\ref{tab:allobjdetails}). Where a \ha-inferred value is available, the predicted \nh\ is consistently lower, and as some of these emission lines are measured rather than upper limits, are likely to be closer to the true value.

\subsection{Intrinsic Nature}\label{sec:nature}

It is difficult at this stage to make definitive statements about the true intrinsic nature of OQQs, but we can speculate on some likely possibilities. A schematic comparison between QSO2s and theories on OQQs can be found in Figure \ref{fig:schematic}. In Section \ref{sec:classcomparison} we discuss how these theories might relate OQQs to various other object classes.

\subsubsection{Young AGN}

A form of changing look AGN \citep[e.g.][]{yang_discovery_2018}, objects belonging to this class have been recently activated such that they heat the torus to produce strong IR signals. However, radiation has not extended far enough to ionise the outer NLR, where the \oiii\ we normally see in QSO2s is produced (middle diagram in Figure \ref{fig:schematic}). For example, \citet{gezari_iPTF_2017} present an example of a `switching-on' AGN that goes from the appearance of a LINER to a Type-I quasar over less than 1 year. No \oiii\ is found in the `before' or `after' spectra, due to \oiii\ following longer term AGN activity than other narrow emission lines, such as \ha, which increase strongly. If OQQs are an obscured analogue to this object, it follows that we are unable to detect other emission lines. They must be observed through the torus, as otherwise we would still observe the broad emission lines produced closer in. If this is the case, and we expect broad emission lines to be present, then detection of these in spectral ranges less subject to obscuration could provide evidence for this explanation \citep[e.g. NIR measurements; ][]{onori_detection_2017}. 
An analogous and opposite population would be those described in \citet{saade_nustar_2022} which have low X-ray luminosity when compared with their strong \oiii\ emission, with one possible explanation being a `switching-off' AGN. They find an example with NLR and nuclear MIR emission but low X-ray luminosity and conclude that this object has recently become inactive. Discovery of this object among their relatively small parent sample implies that AGN caught `mid-change' may not be extremely rare; we might expect `switching-on' AGN to be found in similar numbers.

\subsubsection{`Cocooned' AGN}\label{sec:cocoon}

\begin{figure*}
	\includegraphics[width=0.95\textwidth]{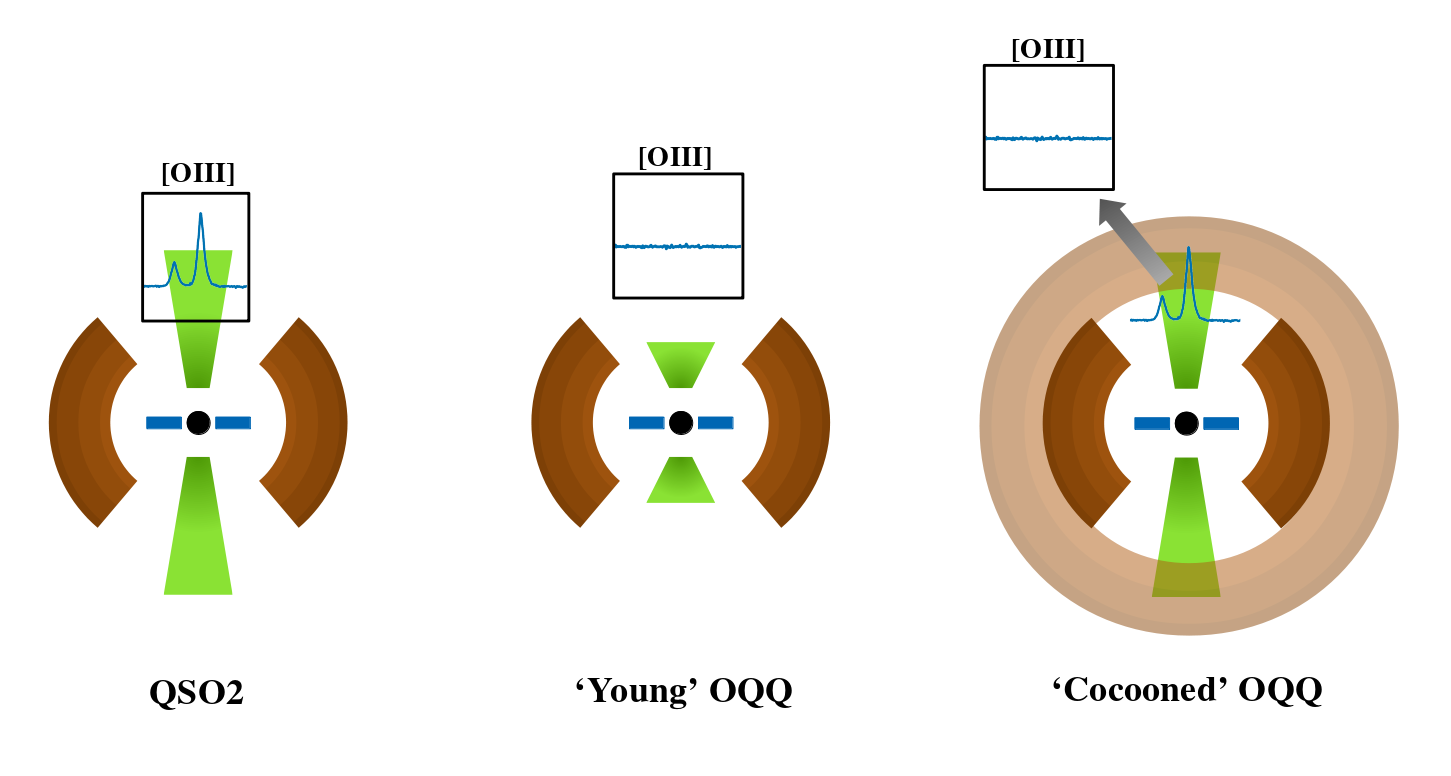}
    \caption{Schematic showing (left) a typical QSO2, producing a spectrum with strong \oiii, (middle) an OQQ that is intrinsically `young', with no \oiii\ produced, and (right) a `cocooned' OQQ, producing typical \oiii\ emission that is obscured by a larger region of material, with the observer positioned to one side of the diagram. Inset boxes show a zoom in on the \oiii\ region of the spectrum - a strong line in the QSO2, but not detected in the two OQQs. Components are as follows: black central SMBH, blue accretion disk, dark brown standard torus, green ionised gas, lighter brown `cocoon'. Composition of dust/obscuring material may be different across various parts of these structures.}
    \label{fig:schematic}
\end{figure*}

This type of OQQ would be more mature than a young AGN, and has similar intrinsic features to a QSO2, but with a `cocoon' of obscuring material preventing either observation of the NLR (right diagram in Figure \ref{fig:schematic}), or its formation entirely. This `cocoon' is likely to be unstable, and as such these objects may only have a short lifetime, suggested by the small number counts found when compared to QSO2s.  

At $z$\,$\approx$\,0.5 for most of our sample, a physical size of 1\,kpc corresponds to an angular size of \SI{0.16}{\arcsecond}, so any circumnuclear dust on sub-kpc scales would certainly be unresolved in most ground-based imaging. If the cocoon is comparable in radius to the dust sublimation radius, it may be even smaller - pc-scale - and thus unresolvable even with instruments such as ALMA. However, high resolution imaging might be expected to show very late stage mergers or other disturbances in the centres of these galaxies, as the source of the `cocoon', despite the `cocoon' itself being unobservable. We searched public archives such as MAST\footnote{{\tt https://archive.stsci.edu/}} and the Virtual Observatory\footnote{{\tt http://www.usvao.org/}} for any high angular resolution imaging of our targets, but these fields have not been observed with \hst, nor with any other large ground-based observatory. In \sdss\ imaging (as well as \wise), the targets do not show any special characteristics, preferred morphologies or orientations, nor obvious disturbances. This appears to support small physical scales for any putative dust cocoons in our sample. However, it must be cautioned that \sdss\ imaging is seeing-limited and not very deep, so for fainter sources, signatures of interaction may be missed.

\subsection{Comparison with other object classes} \label{sec:classcomparison}

Although the idea of optically quiescent AGN is not new, there are no close counterparts of our selected sample, at least in the local universe. In this section, we briefly compare and contrast other related samples with the OQQs.

\subsubsection{ULIRGs}\label{sec:ulirgs}
Our selected targets have total MIR powers within or close to the ULIRG regime. ULIRGs tend to be hosted in chaotic appearing systems, with little clear structure and sometimes lack important optical emission lines, but are not usually completely devoid of strong emission signatures. For example, Mrk\,231 has much stronger \ha\ than \oiii$\lambda$5007 emission \citep{lacy_galactic_1982, malkan_emission_2017}. Several local ULIRGs are also considered as being \lq buried\rq\ AGN obscured along all lines of sight \citep[e.g. ][]{imanishi_strong_2001, oyabu_akari_2011}. These are typically associated with sources that have LINER-like optical spectra with clear emission lines \citep{imanishi_spitzer_2007}.

As discussed in \S\,\ref{sec:emissionlineresults}, 15 of the 28 sources for which redshifted \ha\ lies within the \sdss\ spectral range show significant \ha\ emission. These sources may be similar to ULIRGs such as Mrk\,231, though the comparison with \sdss-selected QSO2s implies that \ha\ also suffers some extinction, as already discussed. The remaining objects clearly differ from local ULIRGs in their lack of \ha. This may be explained by correspondingly higher NLR dust covering factors in the \lq cocooned AGN\rq\ scenario. It could also be a sign that the dust distribution in the galaxy is less centrally located, additionally obscuring \ha\ from sources other than AGN ionisation.

Compact Obscured Nuclei (CONs) are found in LIRGs and ULIRGs \citep[e.g. ][]{falstad_con-quest_2021}. They are dense, IR bright cores found at the centre of an optically normal galaxy. The strength of the IR emission indicates a heavily obscured nucleus, with either a compact starburst or AGN powering the luminosity. One intensively studied example is NGC 4418 \citep{costagliola_high-resolution_2013, asmus_subarcsecond_2014, ohyama_dusty_2019}, which appears optically as an Sa type galaxy, and is part of an interacting pair. Extreme visual obscuration is demonstrated by a very deep silicate absorption feature in the MIR, and high resolution imagery shows a small, optically thick core. \citet{ohyama_dusty_2019} find that a dusty wind and compact starburst activity are present in the core, but that an additional energy source in the core is still required to explain the observations; it is still currently unknown whether this is an AGN or an extremely compact starburst. In a study with \nustar, \citet{ricci_hard_2021} do not detect it and find that a SF model fits the data better, implying either no AGN or extremely thick column densities. 

CONs are not very well understood, and there are not many examples. \citet{falstad_con-quest_2021} search LIRGs and ULIRGs for compact sources of HCN vibrational emission, which are only found in extreme environments. They find that ~40\% of ULIRGs and ~20\% of LIRGs contain a CON, noting that their sample size is not large, and that the selection in the FIR does introduce some biases. This indicates that CONs are relatively common in (U)LIRGs; however, (U)LIRGs themselves are not extremely common.

OQQs are also bright in the IR, but are at larger distances than these prototypical CONs tend to be. Without high resolution imaging, it is impossible to definitively say whether the OQQs are equally compact. Unlike NGC 4418, OQQs appear to be small red galaxies with no obvious signs of interaction.

\subsubsection{Weak Line and Lineless Quasars}

At higher redshifts, several studies have been made of `Weak Line Quasars' \citep[WLQs; e.g.][]{meusinger_large_2014}. Similarly to OQQs, these appear as normal quasars in terms of continuum emission, but show weak or absent emission lines. Examination of WLQs ($z \lessapprox 2.5$; OQQs: $z \lessapprox 1.0$) with \chandra\ snapshots and multi-wavelength analysis from \citet{wu_x-ray_2012} and \citet{luo_x-ray_2015} find spectra that rise in the optical blue end which are generally lacking in stellar absorption lines, in contrast to the majority of the OQQs. They tend to have weak but detectable X-ray emission. At higher redshift, \citet{shemmer_weak_2010} examine the region around the H$\beta$ and \oiii\ lines of two WLQs at $z \approx 3.5$, finding extremely weak H$\beta$ and no detectable \oiii. \citet{wu_x-ray_2012} show results consistent with a weak, or `anaemic' BLR producing weaker emission lines, and \citet{luo_x-ray_2015} suggest that a `puffed up' inner accretion disc could be preventing ionisation of the BLR. WLQs are not selected or restricted in terms of MIR colour or luminosity, which is a key factor in our OQQ selection. Where possible, spectroscopic measurements of more emission lines, along with targeted X-ray observations, could shed light on the link between low redshift OQQs and higher redshift WLQs.

\citet{laor_cold_2011} suggest a cold accretion disc could produce `Lineless Quasars' - if the accretion rate of the AGN is low, then no ionisation can occur, and no emission lines will be observed. They argue that in the case of a cold disc, the ionising component is the X-ray power law alone. These may cross over with WLQs if the X-ray luminosity is high enough, but will appear lineless if $L_X \ll L_{bol}$. Considering the underluminous X-ray results of OQQ~J0751+4028 (see Section \ref{sec:J0751xray}) this may be one reasonable explanation for an intrinsic lack of lines, if that is the case. The optical-NUV-FUV continuum would be expected to be blue in shape, and at this time the wavelength coverage of OQQs does not extend far enough to further investigate this possibility.

\subsubsection{Low X-ray Scattering Fraction AGN}

An interesting population of AGN selected from hard X-ray surveys has been identified by \citet{ueda_suzaku_2007}. These sources show apparently low X-ray scattering fractions with scattering fractions at least an order of magnitude lower than observed in typical Seyferts, and weak \oiii\ as expected from a geometrically thick torus \citep{ueda_oiii_2015}. Interstellar dust reddening is also likely to play a role in depressing the scattered flux, as evidenced from the fact that a significant fraction appear to be hosted in edge-on host galaxies \citep{honig_dust-parallax_2014}.

\citet{gupta_bat_2021} present a detailed study on 386 \bat\ AGN, looking at the relationship between scattering fraction and various AGN properties. They find a significant correlation between scattering fraction and ratio of \oiii\ to X-ray luminosity - i.e. low soft X-ray scattering fraction correlates with weak \oiii. In the context of OQQs, it follows that if the \oiii\ emission is intrinsically not present (as in the `Young' AGN theory), there may also be less scattering of soft X-rays. However, if the obscuring material in the `cocoon' is sufficiently thick to absorb soft X-rays, we may not be able to distinguish between the theories in this way. In the same paper they find that low scattering fraction AGN are also likely to have higher intrinsic column densities, compounding the problem of selecting obscured AGN via optical methods.

Since our sample is MIR--selected, it may be better suited to search for Compton-thick counterparts to low X-ray scattering AGN. Detailed comparison of the OQQs to this sample will require more X-ray observations than are available so far. 

\subsubsection{X-ray Bright Optically Normal Galaxies (XBONGs)}
The class of objects known as XBONGs are X-ray selected AGN that are hosted in optically-normal galaxies lacking (optical) AGN signatures. This lack has been attributed to extinction along all lines of sight, or due to dilution by host galaxy light in a substantial fraction. However, as opposed to our targets, XBONGs tend to generally possess lower Seyfert-like luminosities \citep{moran_hidden_2002, comastri_nature_2002, civano_hellas2xmm_2007, caccianiga_elusive_2007, smith_infrared_2014}. In such cases, it can be easier for the host galaxy to dominate over the AGN and dilute AGN optical signatures. This becomes much harder for the quasar-like luminosities for our targets, and we do not believe dilution to be important for the sample presented herein. 

\begin{figure*}
	\includegraphics[width=\textwidth]{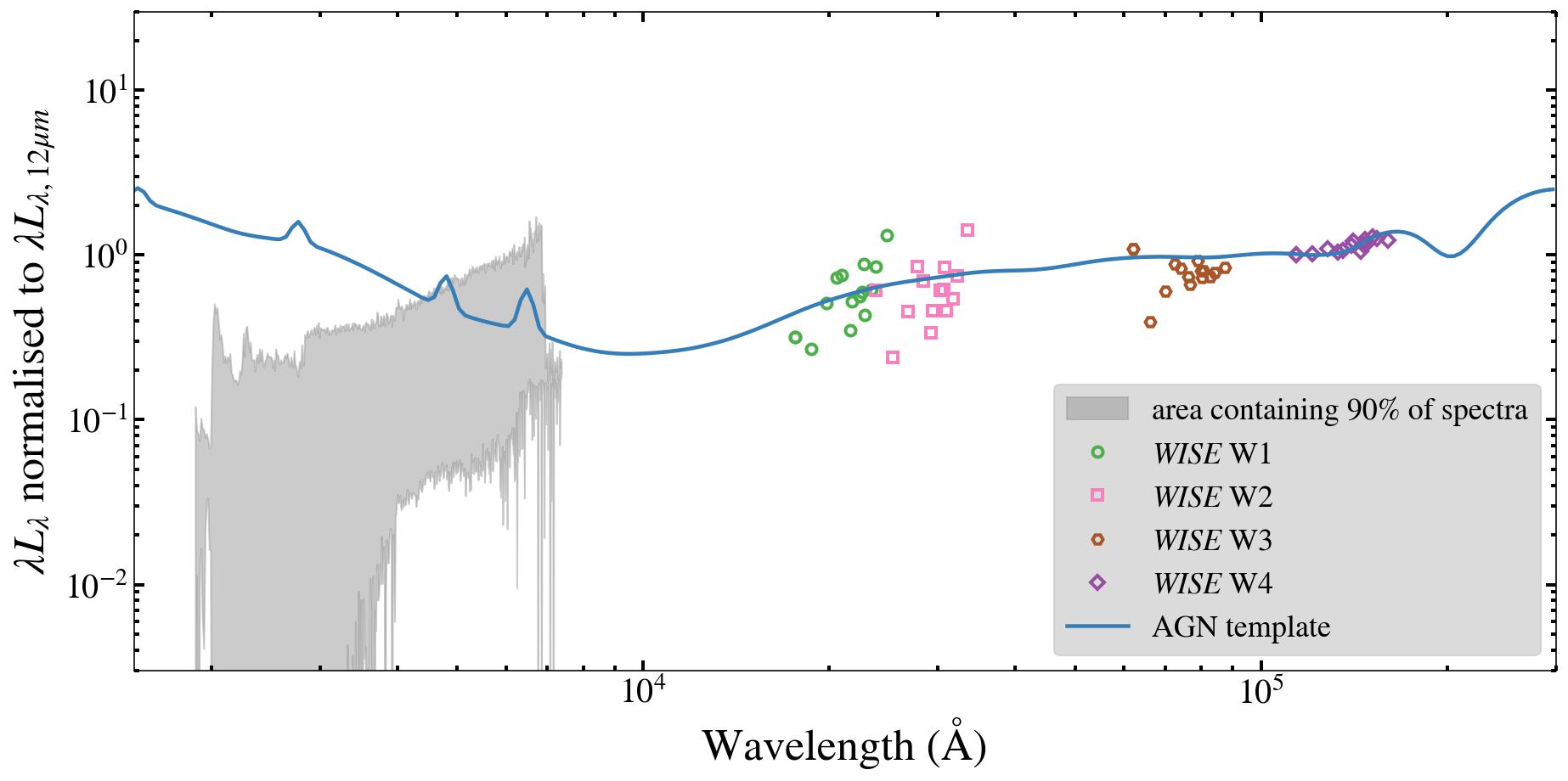}
    \caption{Comparison between \sdss\ spectra (thin grey lines), \wise\ derived luminosities (coloured points) and the AGN spectral template (solid blue line) from \citet{assef_low-resolution_2010}.}
    \label{fig:spectempcomp}
\end{figure*}

In order to demonstrate this, in Fig. \ref{fig:spectempcomp} we compare the OQQ broadband SEDs with the low resolution broadband AGN template compiled by \citet{assef_low-resolution_2010}. The SEDs have been normalised at \SI{12}{\micro\metre} for comparison. All wavelengths referred to here are in rest-frame. The median optical--to--MIR flux ratio for our sample is 0.31 when measured at \SI{5500}{\angstrom} and \SI{12}{\micro\metre}. The template AGN SED clearly lies {\em above} most of our sample in terms of median optical normalised fluxes, with an optical--to--MIR ratio of 0.40. This implies that host galaxy dilution alone is unlikely to be a major effect, if our sample harbours typical central AGN. 
Not all XBONG classifications are considered to be a result of dilution, and XBONGs that are good candidates to be fully covered (\citealt{maiolino_elusive_2003}) may constitute the lower luminosity end of our selected sample.

\subsubsection{High Intrinsic Accretion Rate AGN}

An anti-correlation between \oiii\ and the relative intensity of Fe {\sc II} has been shown in analyses of emission-line and accretion properties of AGN \citep[Eigenvector 1; e.g.][]{boroson_emission-line_1992, shen_diversity_2014}. This relationship can be explained by differences in Eddington ratio - an intrinsically high accretion rate could be associated with a \textit{low} \oiii\ luminosity. To assess the plausibility of the scenario for statistically significant numbers of the OQQ population, accurate measurements of the black hole (BH) masses would be needed; while this is out of the scope of this paper, analysis of NIR broad lines that may be less obscured \citep[e.g. NIR measurements; ][]{onori_detection_2017a} could provide the necessary information. Results from OQQ~J0751+4028 do not support this \citep{greenwell_candidate_2021, greenwell_xmm_2022}, as it is not luminous enough to imply a sufficiently high Eddington ratio.

%\subsubsection{Green Pea Galaxies}

\subsubsection{Other IR-selected Quasars}

Using the \wise-\wise\ colour plane, as done here for OQQs, \citet{hviding_characterizing_2018} focus on a different region of the plane, choosing targets more likely to be heavily obscured. They also make no selection based on emission lines, and the majority of their sources appear more optically typical than the OQQs. However, while there is no direct overlap between these samples, they are likely to be close cousins with unusually heavy obscuration instead of unusually distributed. Further investigation of the targets with insufficient spectral lines or features to assign a spectroscopic redshift leads \citet{hviding_characterizing_2018} to hypothesise that a subset of these may be powerful AGN with emission lines attenuated by thick line of sight obscuration.

Extremely red quasars (ERQs) are quasars that show an unusually high IR to optical ratio \citep[e.g.][]{banerji_heavily_2015, glikman_wise_2022}. \citet{ross_extremely_2015} select these objects based on a comparison between \sdss\ and \wise\ magnitudes, producing a sample of objects up to $z\sim$4. An interesting subset of these, at $z\sim$2-3, show extremely broad emission lines which are not easily explained by current models. Some of these ERQs could be a Type 1 analogue to \lq cocooned\rq\ OQQs - i.e. sources with a high covering factor (not the complete $4\pi$ steradians) and possibly obscuring material on large scales, but that retain some (unobscured but possibly reddened) line-of-sight to their nucleus; therefore presenting with broad AGN emission lines. Spectropolarimetry results from Type 2 ERQs \citep{alexandroff_spectropolarimetry_2018} indicate that the visible emission lines are due to reflection from a dusty disc wind rather than a direct line-of-sight. A potential interpretation is that these are similar to \lq cocooned\rq\ OQQs, but with a small opening that allows the opportunity for reflected emission to be visible. Furthermore, an attempt at OQQ selection in large surveys based on photometry alone or low-quality spectroscopic data may include a number of ERQs, therefore it is important to understand their physical differences and how they might be distinguished.

Another interesting population found with extremely high infrared luminosities are Dust Obscured Galaxies (DOGs). These objects are selected based on an excess of \SI{24}{\micro\metre} (from \spitzer; \citealp{dey_significant_2008}) or \SI{22}{\micro\metre} (from \wise; \citealp{toba_hyper_2015}) emission compared to optical, and are thus selected to be underluminous in the rest-frame UV. They often show signs of intense star formation or obscured AGN activity \citep{dey_significant_2008, toba_far_2017}, with MIR colours indicating AGN presence in a large number of them, particularly at higher luminosities \citep{toba_search_2016, toba_far_2017}. The high luminosity along with dust obscuration seem to place them into the high-growth post-merger phase of the galaxy merger scenario for AGN-host galaxy growth \citep{hopkins_cosmological_2008}. Their selection process results in a subset of DOGs that lack broad emission lines, and thus may be closer to OQQs than Type 1 ERQs. However, evidence of star formation and the presence of strong narrow lines shows that they are not an identical population. Spectropolarimetry results may suggest an evolutionary connection between ERQs, Hot DOGs, and other reddened quasars \citep{assef_imaging_2022}.

\citet{glikman_first-2mass_2012} select candidate red quasars based on \twomass\ (IR) and \first\ (radio) data. A detailed examination of the radio properties of OQQs is outside the scope of this paper, but we searched the \first\ catalogue \citep{becker_first_1995} and found 7 \first\ counterparts to OQQ candidates, of which four are brighter than 10mJy at 1.4~GHz. The OQQs are therefore not consistently radio bright or dim, and unlikely to be similar, population-wise, to radio-selected objects.

\subsubsection{Blazars}\label{sec:blazars}

One potential source of contamination in this sample is likely to be blazars - the chance viewing of AGN head on to a relativistic jet \citep{landt_physical_2004}, producing a bright, variable source, strong in radio emission. These are known to be very bright objects, with some exhibiting very flat spectra. As such, there may be a number that pass the criteria specified in Section \ref{sec:sampleselection}, but that are not relevant objects for this paper. It would be useful therefore to be able to estimate the level of such contaminants in our sample, and if possible to remove some in order to improve the reliability of the sample.

\begin{figure}
	\includegraphics[width=\columnwidth]{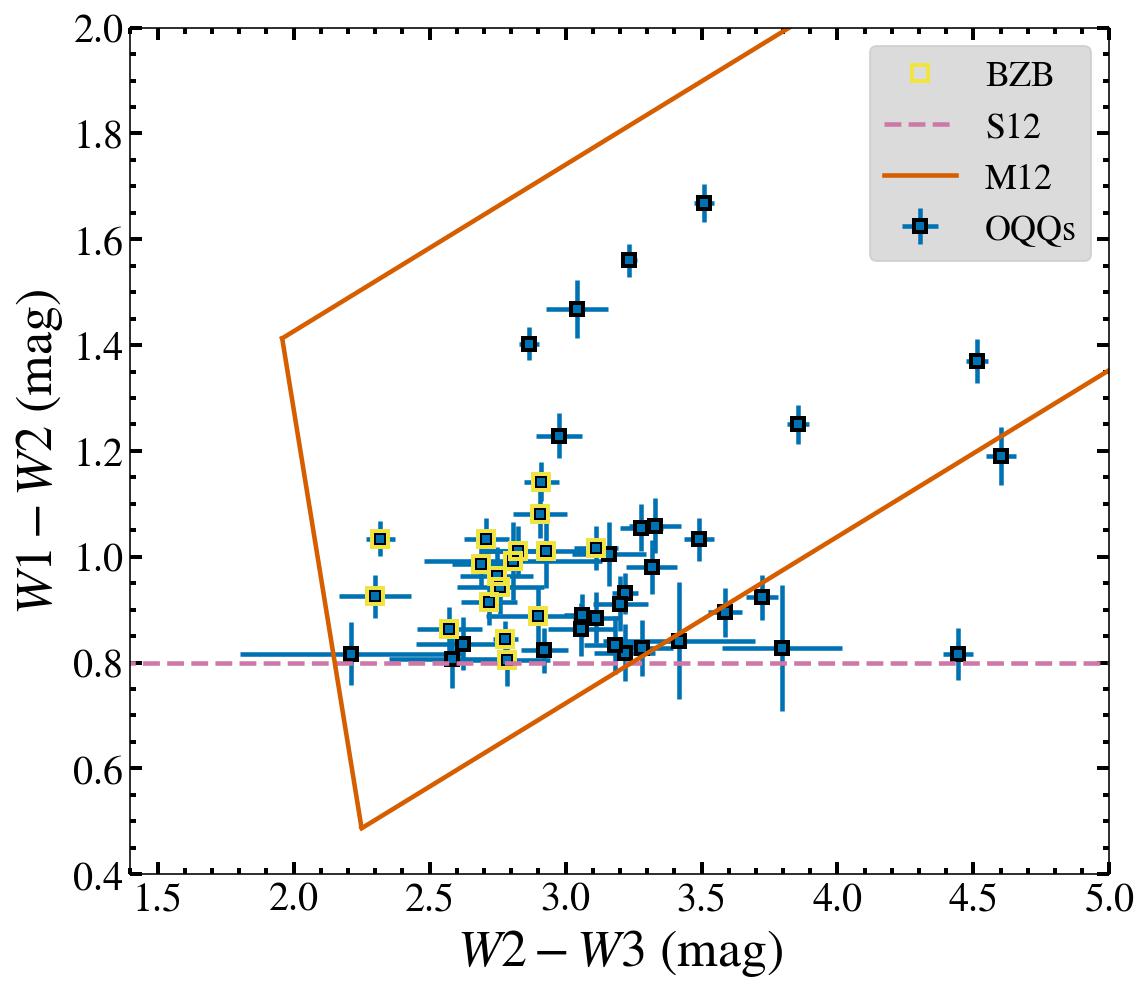}
    \caption{\wise\ colour-colour diagram (as in selection; see Figure \ref{fig:wisecolours}) showing the categorisation of the OQQs. OQQs that pass in all three planes of \citet{massaro_wise_2012} are outlined with yellow squares.}
    \label{fig:blazarstripbzb}
\end{figure}

\citet{dabrusco_wise_2014} created a catalogue of $\gamma$-ray emitting blazars, assembled from \wise\ sources. Comparison of overlapping sources from this catalogue (\textit{WIBRaLS}) with our OQQ sample, and with respect to various parameters, could help identify the unwanted objects. \citet{massaro_wise_2012} identified the \wise\ Gamma-ray Strip (WGS), which uses IR colour-colour diagrams to distinguish possible blazars. They divide these into two sub-categories - BZBs (BL Lac objects, with featureless optical spectra - galactic absorption lines or weak emission lines only), and BZQs (flat spectrum radio quasars, where the optical spectrum should show broad emission lines). According to these definitions, the most likely category for any contaminant objects will be BZB. \citet{massaro_wise_2012} quantify the likelihood of a source being within the WGS, based on its location on three \wise\ colour-colour diagrams, and the uncertainties on those colours. Figure \ref{fig:blazarstripbzb} shows the application of this method to our objects. This is not considered a sufficiently reliable classification to remove any sources from the OQQ list, merely to assess the level of contamination from BZB sources.

Furthermore, given the flat spectrum characteristic of blazars, they tend to have weak values of D4000. Figure \ref{fig:breakdata} shows that the OOQs tend to show significant breaks, and this can be seen in the individual optical spectra (Figure \ref{app:opticalspectra}). Hence a visual check of the spectra substantially reduces the likelihood of blazar contamination (Section \ref{app:vischeck}).

One property common to all blazar types is that they exhibit high, irregular variability across the whole spectrum. As such, this may be a useful way to distinguish likely blazars from OQQs.

After the completion of the main \wise\ mission, the \neowise\ project \citep{mainzer_preliminary_2011} continues to scan the sky, doing so approximately every six months (depending on sky location) with the \textit{W1} and \textit{W2} bands (\textit{W3} and \textit{W4} cannot operate without cryogenic cooling, and this was exhausted during the main phase). The primary science aim of this phase is the study of near Earth objects. However as the single exposure images and source magnitudes are available from the \neowise\ database\footnote{\url{http://wise2.ipac.caltech.edu/docs/release/neowise/}}, it provides a measure of the multi-year timescale infra-red variability.

\begin{figure}
	\includegraphics[width=\columnwidth]{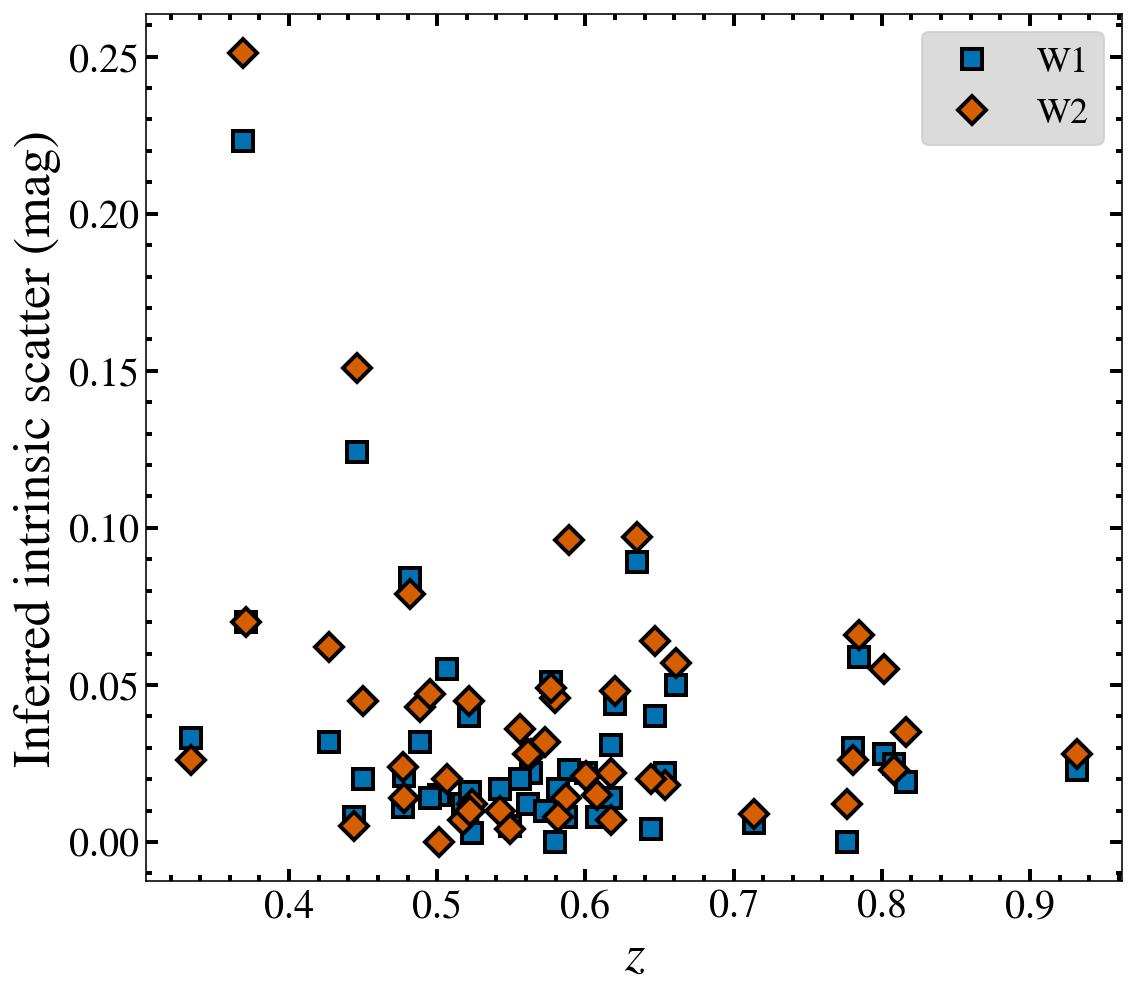}
    \caption{Intrinsic spread added to force the \wise\ data to be consistent with an unvarying source - higher values indicate that the source is more variable, and less likely to be stable.}
    \label{fig:wisevarestimate}
\end{figure}

\begin{figure}
	\includegraphics[width=\columnwidth]{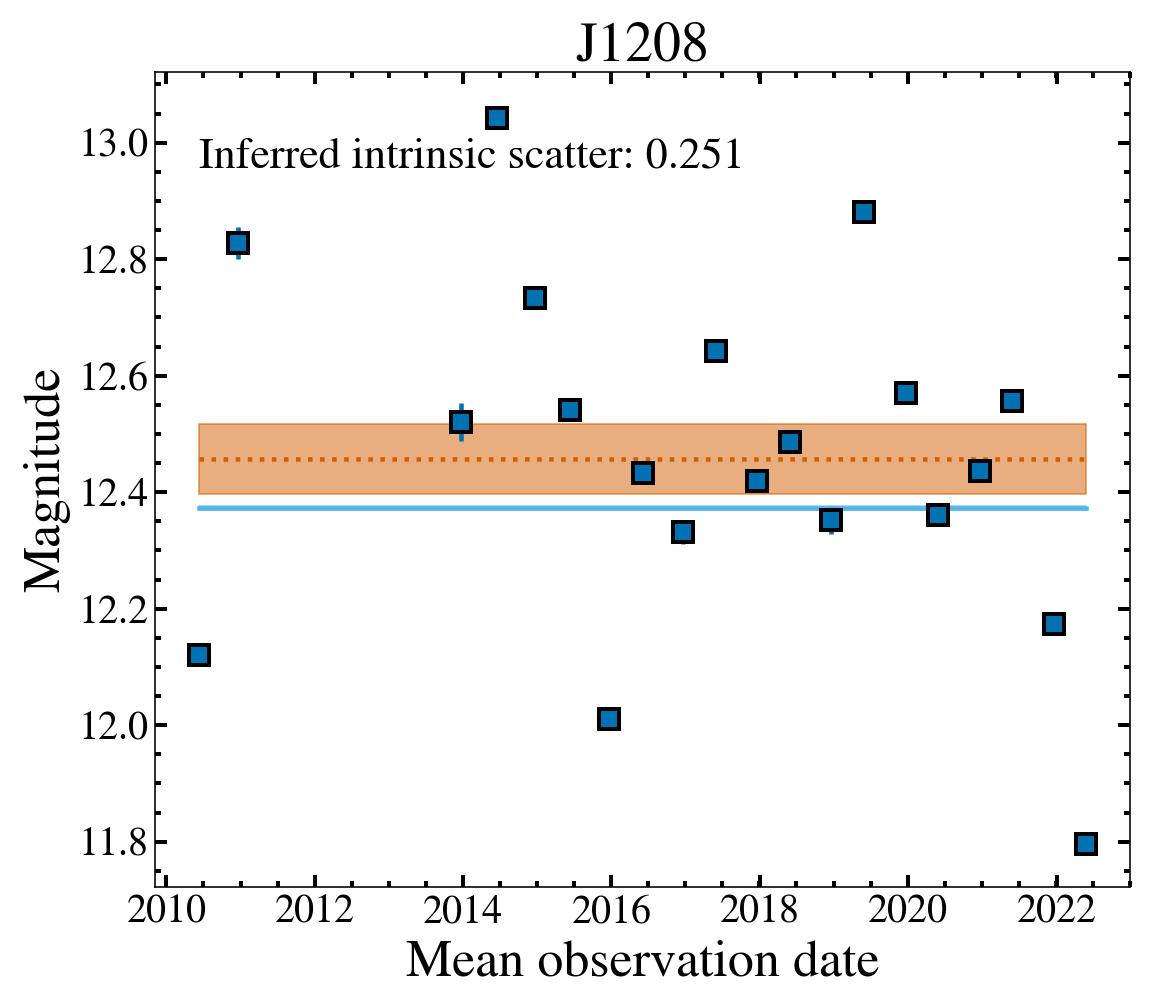}
    \caption{\textit{W2} variability of the OQQ showing significant variation ($>$0.1~mag). It appears to show random variability, although the large time steps in available \wise\ data make statistical assessment of the variability difficult.}
    \label{fig:wisevarhigh}
\end{figure}

To approximately quantify the level of variability in each source, the data from each \wise\ measurement epoch was averaged, then these combined points fit to a time-invariant line. $\chi^2$ was used to assess whether this invariant state could reasonably explain the observed variation in  \wise\ detections\footnote{The \textit{W3} (\SI{12}{\micro\metre}) band magnitudes were chosen as most representative of the region of interest. A more detailed investigation would examine the effects with luminosities rather than magnitudes.}. If it could not, then intrinsic uncertainty was added to the points. The size of this intrinsic uncertainty was used as an estimate of the variability of the source, and Figure~\ref{fig:wisevarestimate} shows the results of this.
Figure \ref{fig:wisevarhigh} shows the fitting with the OQQ showing highest variability, OQQ~J1208+1159. This source is flagged as a WGS BZB (see Figure~\ref{fig:blazarstripbzb}), and is detected in \first\ \citep{becker_first_1995} as a compact source with integrated flux \SI{67.31}{mJy}. This combined with the MIR variability suggests that it may be a blazar.

The number of objects found that are above the threshold for being within the WGS BZB is 17 out of 47 OQQ. The number of objects with matches in the WIBRaLS table is 8 out of 47 OQQs - these are all also above the BZB threshold. The number of objects with high variability is currently difficult to quantify without a comparison threshold. The QSO2 variability is slightly lower on average (mean 0.031/median 0.020 for OQQ, mean 0.026, median 0.017, for QSO2s) and includes no very high variability objects.

If we take the lower limit of high variability as 0.1, then there are two OQQs above this threshold, both of which are likely BZBs. Of the 47 OQQs below the threshold, 36\% are likely BZBs. In comparison, 57 QSO2s (out of 1,990; 2.9\%) pass the BZB threshold overall. An examination of the variability of a large sample of non-AGN and various types of AGN would be illuminating, but outside the scope of this work. Due to the high uncertainties on these methods, it is difficult to fully quantify the number of BL Lac objects that could be interfering in the results here: anything from 3\% to 25\% may be possible.

\subsection{Evolutionary Context}

It is important to consider the cosmological situations where we might find OQQs. Many works have examined the life cycle of galaxies and AGN during mergers \citep[e.g. ][]{dimatteo_energy_2005, springel_modelling_2005, hopkins_unified_2006}. As the two galaxies interact, the interstellar medium from both comes together, and the resultant enhanced density of matter leads to increased star formation, and (as the bulk of the merging galaxy reaches the nucleus of its pair) we may see increased AGN activity. This brightening starts as the AGN is still enshrouded in dust and gas from the merger, but as the emission increases in intensity, energy and matter feedback starts to clear out the obscuring matter. Our OQQ sample objects, for the most part, do not show signs of high SFR as may be expected if they lie in this part of the sequence. Thus, OQQs do not obviously fit the standard paradigm for AGN and host galaxy evolution.

\citet{springel_modelling_2005} show that in a modelled merger of two disc galaxies, including AGN feedback and BH growth, the peak of SFR precedes the black hole accretion rate (BHAR) peak by a small amount of time. SFR begins to drop before the BHAR, so there may be a short period of time where the SFR is well below peak, but BHAR is high. As BHAR is proportional to AGN luminosity, the OQQ objects might be objects in this short transitional phase. We find fewer OQQs compared to QSO2s, suggesting that the OQQ duty cycle is shorter than the duty cycle of Type 2 quasars. Interestingly, the difference in median redshift between these two samples (e.g., Fig.\,\ref{fig:zL12dist}) implies that there may be genuine evolutionary differences between their populations, with OQQs peaking earlier than QSO2s. However, incompleteness issues could impact both source samples at the upper redshift end, and the number of nearby OQQs may be artificially low simply due to easier detectability of faint \oiii.

The two galaxies are coalescing at this point so would not necessarily be clearly identifiable as advanced mergers, especially with the images currently available. Additional high resolution imagery could be valuable in identifying any late stage mergers among the sample. For the OQQs to be objects existing in this brief period of time, we may expect strong outflows to be visible, if the SFR is being quenched by AGN activity. These will manifest as blueshifted absorption lines, or broadened and asymmetric emission lines at the systemic velocity of the source. With the current spectra this is impossible to either confirm or rule out, but measurements of select likely candidates with more sensitive instruments could provide more insight into these scenarios.

Simulations by \citet{blecha_power_2018} suggest that MIR selection of AGN in mergers may miss AGN, particularly low apparent luminosity and very late stage mergers. This work selects for high IR luminosity objects in order to produce a more reliable candidate list; lower luminosity OQQs may be missed from MIR selection due to (a) selection being less complete in general for less bright objects, (b) as shown in \citeauthor{blecha_power_2018} if OQQs are late stage mergers.

The absence of obvious merger signatures may instead point towards high mass ratio mergers, with the smaller galaxy not massive enough to significantly disrupt the morphology of the larger primary, nor massive enough to trigger substantial star formation throughout the body of the primary. However, an efficient mechanism to channel the gas from the donor galaxy to the central SMBH would be required in this case. Alternatively, whether {\em secular} processes such as efficient gas flows can trigger AGN activity without simultaneous star formation remains unclear. 

Alternatively, the \lq Young AGN\rq\ scenario presents a different evolutionary perspective. If the triggering event is recent enough, the AGN could be switching on, mimicking the phase described by \citet{schawinski_active_2015} as \lq Optically Elusive AGN\rq. They show that there could be a period of $\sim$10$^4$ years in which the AGN is visible in X-rays but has not yet photoionized the extended area that will later become the NLR. As the dusty torus (the source of reprocessed IR emission, and the IR colour that we select for in this work) is closer to the central engine than the NLR, we expect that sign of AGN life to appear prior to any narrow optical emission lines.

\citet{schawinski_active_2015} calculated the full duty cycle of an AGN as $\sim$10$^5$ years, and the \textit{optically elusive} phase as $\sim$10$^4$ years. If we expect the IR bright OQQ phase to be some fraction of the optically elusive phase, this puts the OQQ lifetime at less than $\sim$10\% of the \lq normal AGN\rq\ duty cycle. The viewing angle of OQQs and QSO2s is likely to be similar - both viewed through the torus, obscuring any broad emission lines - so assuming they have similar covering factors, we would expect to see them in the same proportions as the full range of AGN types: approximately 10:1. We find $\sim$1000 QSO2s and 47 OQQs: approximately 20:1; encouragingly close, but somewhat fewer OQQs than expected, especially if we consider that some OQQs may be \lq cocooned\rq\ rather than \lq young\rq. As stated above, however, the 10:1 ratio is a \textit{lower limit}, based on the full \lq optically elusive\rq\ timescale. If every OQQ is a \lq young AGN\rq\ our results are consistent with the IR bright stage being approximately two thirds of the \lq optically elusive\rq\ phase. These are merely crude estimates that do not fully account for selection and modelling biases, much of which remain unknown.

An investigation of the morphologies of OQQs could provide some interesting insight into their relationship with their host galaxies; for example, \citet{rigby_why_2006} found from an examination of the morphologies of X-ray--selected optically-dull AGN that such sources exhibited a wider range of axis ratios (major-minor axis, assuming an elliptical shape) as compared to optically-active AGN which showed only very round axis ratios. They conclude that extranuclear dust within the host galaxy is responsible for an optically-dull nature. \sdss\ photometry is severely limited for this purpose and is unlikely to be able to accurately resolve the size or axis ratio of the host galaxy except in unusual cases. Future surveys, such as Euclid \citep{euclid_euclid_2022}, will greatly improve on this.

\section{Summary}

By selecting MIR--luminous sources with red \wise\ colours and a marked absence of optical emission lines, we have presented a new sample of quasars that are candidates for being enshrouded within dust cocoons. The overall sample shows broadband continuum properties very similar to those of \sdss-selected QSO2s, but with an \oiii$\lambda$5007 emission line suppression factor of $\gtsim$250 (see Section~\ref{sec:emissionlineresults}). We show that host galaxy dilution is unlikely to be a major effect in most cases.  

The idea of fully enshrouded AGN is far from new. Various models suggest that the bulk of AGN growth occurs in highly obscured phases \citep[e.g. ][]{fabian_mass_1999,hopkins_unified_2006}. The shape of the cosmic X-ray background spectrum also requires similar numbers of Compton-thick (\nh $>$ $10^{24}$ cm$^{-2}$) obscured AGN compared to unobscured. \citet{ananna_accretion_2019} calculate the required proportion at $z=1.0$ as 56\% $\pm$ 7\% (50\% $\pm$ 9\% at $z=0.1$). A number of dedicated legacy surveys with \nustar\ that are  selected in the infrared are finding CT fractions $\gtrsim$30\% within $\lesssim$200\,Mpc (e.g., the \nustar\ Local AGN NH Distribution  Survey -- NuLANDS, Boorman et al. in prep., and the [Ne V] survey; \citealp{annuar_nustar_2020}). In any evolutionary paradigm, this would naturally include fully obscured sources in the stochastic competition between gas feeding from large scales and AGN feedback, especially if there is a delay between the two competing processes (noting that NuLANDS  currently includes optical classification as part of the sample definition; the preliminary infrared selection should still include such fully-covered  objects). It may also imply the existence of obscured objects that do not occur as part of the merger-driven growth phase, but instead represent a separate path.

The receding torus model also suggests that the covering factor of AGN decreases with luminosity \citep[e.g. ][]{lawrence_relative_1991} so fully covered AGN may be more likely to occur at low AGN luminosities. This issue is far from settled, however, with conflicting results from a variety of studies in different bands \citep[e.g. ][]{glikman_first_2007, lawrence_misaligned_2010, toba_luminosity_2014, assef_half_2015}. An element of bias can be introduced by requiring that optical emission lines be present for secure source redshift and type identification - any sample based on this selection criteria may be missing objects with some emission lines blocked, either completely or to a level below measurable. The strength of our selection criteria is to instead rely upon the reprocessed thermal spectrum (i.e. MIR colours) and luminosities, allowing us to search for objects in which potential 4$\pi$ or near 4$\pi$ covering completely extinguishes the optical lines. 

\begin{itemize}
    \item We show that unknown AGN with unusual properties can be selected with a combination of multi-wavelength selection techniques, finding 47 \lq Optically Quiescent\rq\ Quasars (OQQs; strong MIR objects with no \oiii\ emission) -- see Section \ref{sec:sampleselection}.
    \item Comparisons with Type 2 quasars (QSO2s; similarly MIR bright, but selected for strong narrow \oiii) show similar underlying properties in e.g. redshift, colour, and luminosity distributions, but fewer numbers of OQQs -- see Section \ref{sec:qso2s}.
    \item The exception to this is a strong deficit between the detected \oiii\ emission from QSO2s and the upper limits placed on the OQQ lines -- see Section \ref{sec:emissionlineresults}.
    \item We consider what physical explanations may produce the observed properties (Section \ref{sec:nature}), principally:
    \begin{itemize}
        \item \lq Cocooned\rq\ AGN - an accreting black hole entirely shrouded in obscuring material.
        \item \lq Young\rq\ AGN - a recently switched on AGN, resulting in contrasting signals originating at different distances from the black hole.
        \item Intrinsically weak lined AGN - some properties of the accretion disc can cause ionisation of narrow lines to be less prominent.
    \end{itemize}
    \item Comparisons with other populations of AGN show that OQQs do not fit into any of these groups, but may be similar to some, depending on their intrinsic nature - e.g., Weak Line Quasars may be similar to OQQs that are caused by unusual accretion activity that results in low ionisation; Compact Obscured Nuclei may be similar to \lq cocooned\rq\ AGN -- see Section \ref{sec:classcomparison}.
\end{itemize}

\section*{Data Availability Statement}

The data underlying this article are publicly available from the \wise\ All-Sky Survey, \sdss\ DR15, and X-ray data from the HEASARC archive.

\section*{Acknowledgments}

We would like to thank the referee for their helpful and constructive feedback.
This research is funded by UKRI. CG received support from a University of Southampton Mayflower studentship and Durham University CEA STFC grants, numbers ST/T000244/1 and ST/X001075/1. PG acknowledges support from STFC and a UGC-UKIERI Thematic partnership (STFC grant number ST/V001000/1). PGB acknowledges financial support from the Czech Science Foundation project No. 22-22643S. \wise\ is a project of Univ. California, Los Angeles, and Jet Propulsion Laboratory (JPL)/California Institute of Technology (Caltech), funded by the National Aeronautics and Space Administration (NASA). Funding for \sdss-III has been provided by the Alfred P. Sloan Foundation, the Participating Institutions, the National Science Foundation, and the U.S. Department of Energy Office of Science. The \sdss-III web site is http://www.sdss3.org/. \twomass\ is a project of the Univ. Massachusetts and the Infrared Processing and Analysis Center/Caltech, funded by NASA and the National Science Foundation. This research has made use of data obtained from the \chandra\ Data Archive and the \chandra\ Source Catalog, and software provided by the \chandra\ X-ray Center (CXC) in the application packages CIAO and Sherpa. This work made use of observations obtained with \xmm, an ESA science mission with instruments and contributions directly funded by ESA Member States and NASA. This research made use of data from the \nustar\ mission, a project led by the California Institute of Technology, managed by the Jet Propulsion Laboratory, and funded by NASA. This research has made use of the \nustar\ Data Analysis Software (NuSTARDAS) jointly developed by the ASI Science Data Center(ASDC, Italy)and the California Institute of Technology(USA). The NASA/IPAC Infrared Science Archive (IRSA) operated by JPL under contract with NASA, and SIMBAD operated at CDS, Strasbourg, France, were the main databases queried. The USNO Image and Catalogue Archive is operated by the U.S. Naval Observatory. This research has made use of data obtained through the High Energy Astrophysics Science Archive Research Center Online Service, provided by the NASA/Goddard Space Flight Center. This research made use of Astropy\footnote{http://www.astropy.org}, a community-developed core Python package for Astronomy \citep{astropy_2013, astropy_2018}. Many thanks also to Taiki Kawamura for his helpful comments and suggestions. 

%%%%%%%%%%%%%%%%%%%%%%%%%%%%%%%%%%%%%%%%%%%%%%%%%%



\bibliographystyle{mnras}
\bibliography{library_bibtex}

%%%%%%%%%%%%%%%%%%%%%%%%%%%%%%%%%%%%%%%%%%%%%%%%%%

%%%%%%%%%%%%%%%%% APPENDICES %%%%%%%%%%%%%%%%%%%%%

\appendix

\section{Source spectra} \label{app:opticalspectra}

\begin{figure*}
	\includegraphics[height=0.9\textheight,keepaspectratio]{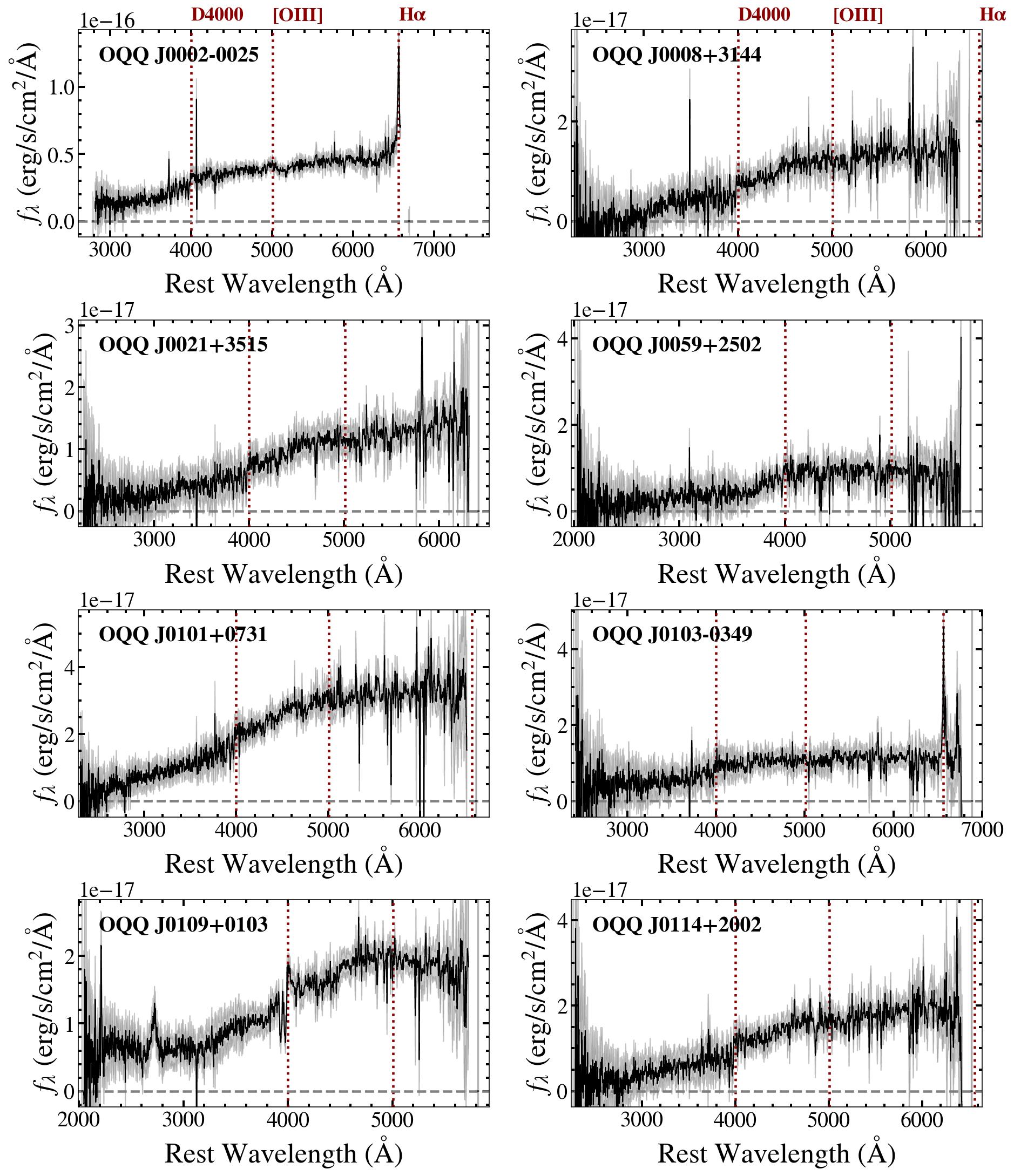}
% 	\vspace*{-12mm}
    \caption{\sdss\ spectra of the first eight OQQs by RA, smoothed to show the continuum shape of each spectrum. Dashed lines in dark red mark key features: the 4000\AA\ break (see Section \ref{sec:d4000}), \oiii, and \ha. Full sample available online.}
    \label{fig:sdss_spectra}
\end{figure*}

This section presents the \sdss\ spectra of a sub-sample of OQQs at rest-frame wavelength in Figure \ref{fig:sdss_spectra}. All spectra are available online.

\section{Spectra Selection}

\subsection{\oiii\ Line Presence Elimination} \label{app:oiiicheck}

\begin{figure*}
	\includegraphics[width=\textwidth]{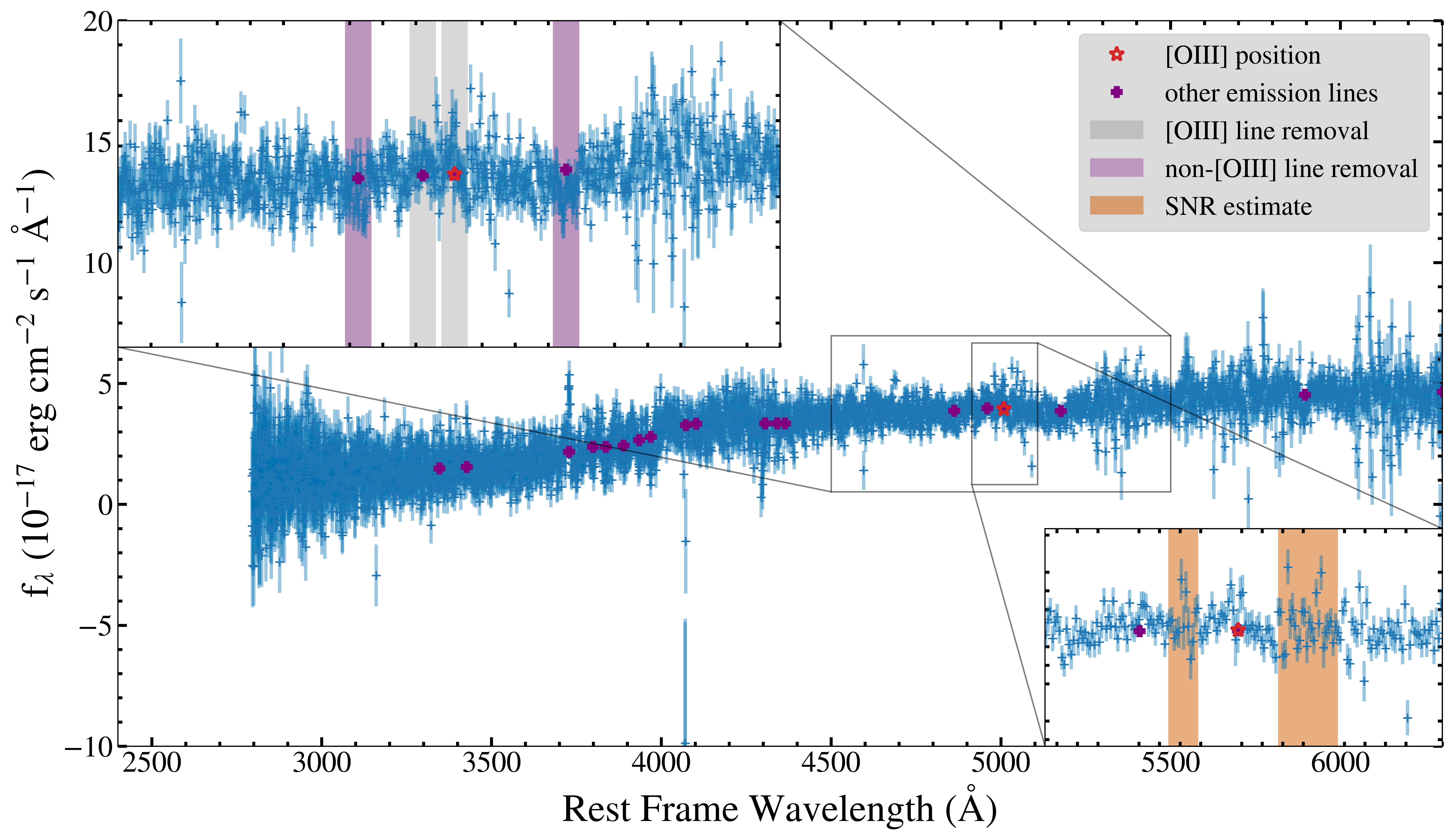}
    \caption{Full spectrum, showing cut out with the region used in the fits. Purple + markers show the possible location of any spectral lines. Red star marker shows the location of \oiii. Vertical shaded regions show emission lines cut for the fit. Bottom right panel shows the region used for SNR calculations. Vertical shaded areas show the SNR estimate regions. Noise is estimated from these and compared to the continuum flux, as described in Section \ref{app:snrcheck}.}
    \label{fig:contfitcut}
\end{figure*}

We start with the primary \sdss\ spectrum for each OQQ candidate, which was cut down to the region around the \oiii\ doublet (rest frame wavelength 4860 \AA\ to 5100 \AA\ - see Figure \ref{fig:contfitcut}). We aim to find upper and lower bounds for a fit to the \oiii\ line, based on a Gaussian on top of a straight line. In \citet{reyes_space_2008}, they used a double Gaussian fit to find the presence of strong \oiii. However, we found that due to the low level of our target \oiii\ lines, the second peak caused the fit to be confused by the noise. We therefore fit only one Gaussian. Next we assess the uncertainty on the line flux, with the aim of checking whether the lower bound is less than zero, and hence whether only an upper bound can be placed on the emission. The continuum can be ignored in this simulation, as it cancels out from the equation. Finally, we select OQQs from the candidate objects that pass all other tests (as outlined in Section \ref{sec:sampleselection}).

% \vspace{5mm}

\begin{figure*}
	\includegraphics[width=\textwidth]{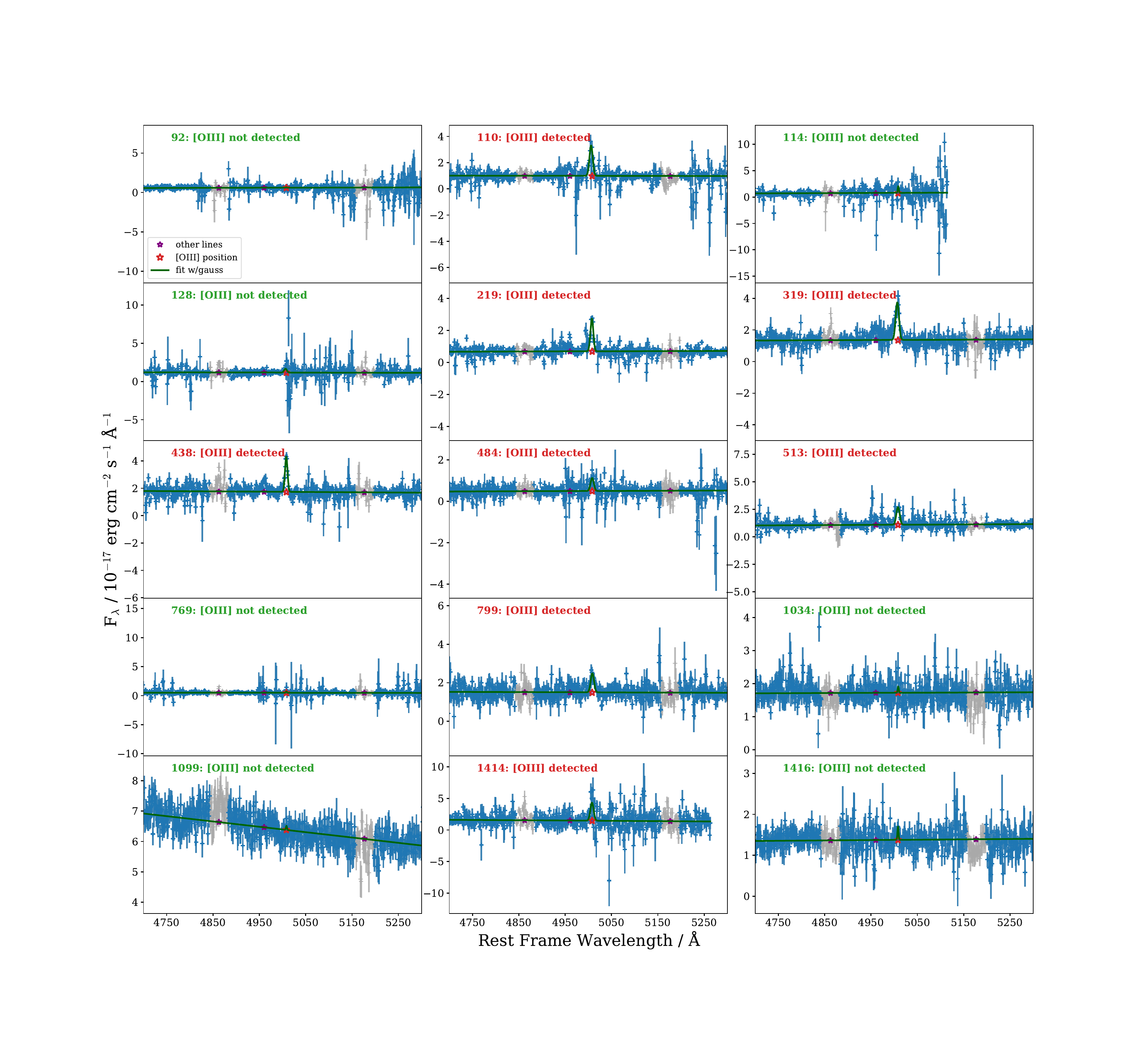}
    \caption{Candidate OQQs, randomly selected from the catalogue at Step 8 (see Table \ref{tab:selection}). Labelled in green (\lq\oiii\ not detected\rq) show the objects that pass this test and move on to visual checks, and those labelled in red (\lq\oiii\ detected\rq) are discarded. Grey sections of the spectra show regions removed for this test to avoid other possible emission lines biasing the fit.}
    \label{fig:oiiicheck_random}
\end{figure*}

\noindent Method:
\begin{enumerate}
    \item cut to local region around \oiii\ line wavelength (see Figure \ref{fig:contfitcut})
    \item Set bounds for all parameters (see Table \ref{tab:params})
    \item Remove regions where any spectral lines other than \oiii\ would be found, if present (see Figure \ref{fig:contfitcut})
    \item Fit function to data ($y=mx+c+gaussian(x_0, \sigma, h)$)
    \item Simulate an \oiii\ line for parameters randomly selected from their uncertainty bounds
    \item Calculate line flux for each simulated line
    \item Look at lower bound on line flux from simulations - if less than zero, we can conclude that the line only has an upper limit (Figure \ref{fig:oiiicheck_random} shows spectra for a selection of \oiii\ detected and non-detected candidates)
    \item Select subset to pass onto the next step.
\end{enumerate}

\begin{table}
	\centering
	\caption{Bounds imposed on fit parameters. Slope and intercept of background continua were allowed to vary freely, whereas parameters of the Gaussian fit to the emission line were restricted to physically likely values.}
	\label{tab:params}
	\begin{tabular}{lll}
\hline
                Parameter &    Symbol &                    Bounds \\
\hline
                    Slope &       $m$ &                      none \\
                Intercept &       $c$ &                      none \\
    Line wavelength / \AA &     $x_0$ &    up to \SI{1000}{\kilo\metre\per\second} from centre \\
          Line FWHM / \AA &  $\sigma$ &   physical min up to \SI{500}{\kilo\metre\per\second} \\
 Line height / flux units &       $h$ &                 max based on data \\
\hline
\end{tabular}
\end{table}

\subsection{Visual check}\label{app:vischeck}

All spectra that passed the automatic check were then inspected by eye, and discarded or flagged based on several criteria. These cuts were done with the aim of making a highly reliable rather than complete sample, so that we are as certain as possible about the relevant nature of each passing candidate. A summary of these results is shown in Table \ref{tab:oiiisummary}.

\begin{enumerate}
    \item Visible \oiii\ - the automatic check sometimes fails to find a significant \oiii\ signal, despite there being a small but clearly apparent peak. We visually examine each spectrum, and based on the region around \oiii\, and the width and height of the visible peak, we discard or keep objects based on whether the \oiii\ seems truly absent.
    \item Noisy data - we discard any candidates where the noise in the region of the \oiii\ line is visibly worse than the flux there, in order to remove any where the \oiii\ is likely to be lost in the noise. The S/N is calculated in detail in a later step (Section \ref{app:snrcheck})
    \item Incorrect class - candidates were selected based on \sdss\ spectroscopic classifications, but rarely an object is misclassified, and we discard any objects where the class appears unreasonable.
    \item Uncertain redshift - candidates were already selected based on \sdss\ spectroscopic redshift showing no warning, but this automatic fit fails occasionally. We discard any objects where it is not clear how the redshift has been assigned.
\end{enumerate}

By design, the OQQ spectra show no \oiii, and as shown in Figure~\ref{fig:lhal12}, \ha\ is also generally low, where measurable. However, some optical spectra show other indicators of AGN presence and/or other interesting features.

\begin{figure*}
	\includegraphics[height=0.9\textheight,keepaspectratio]{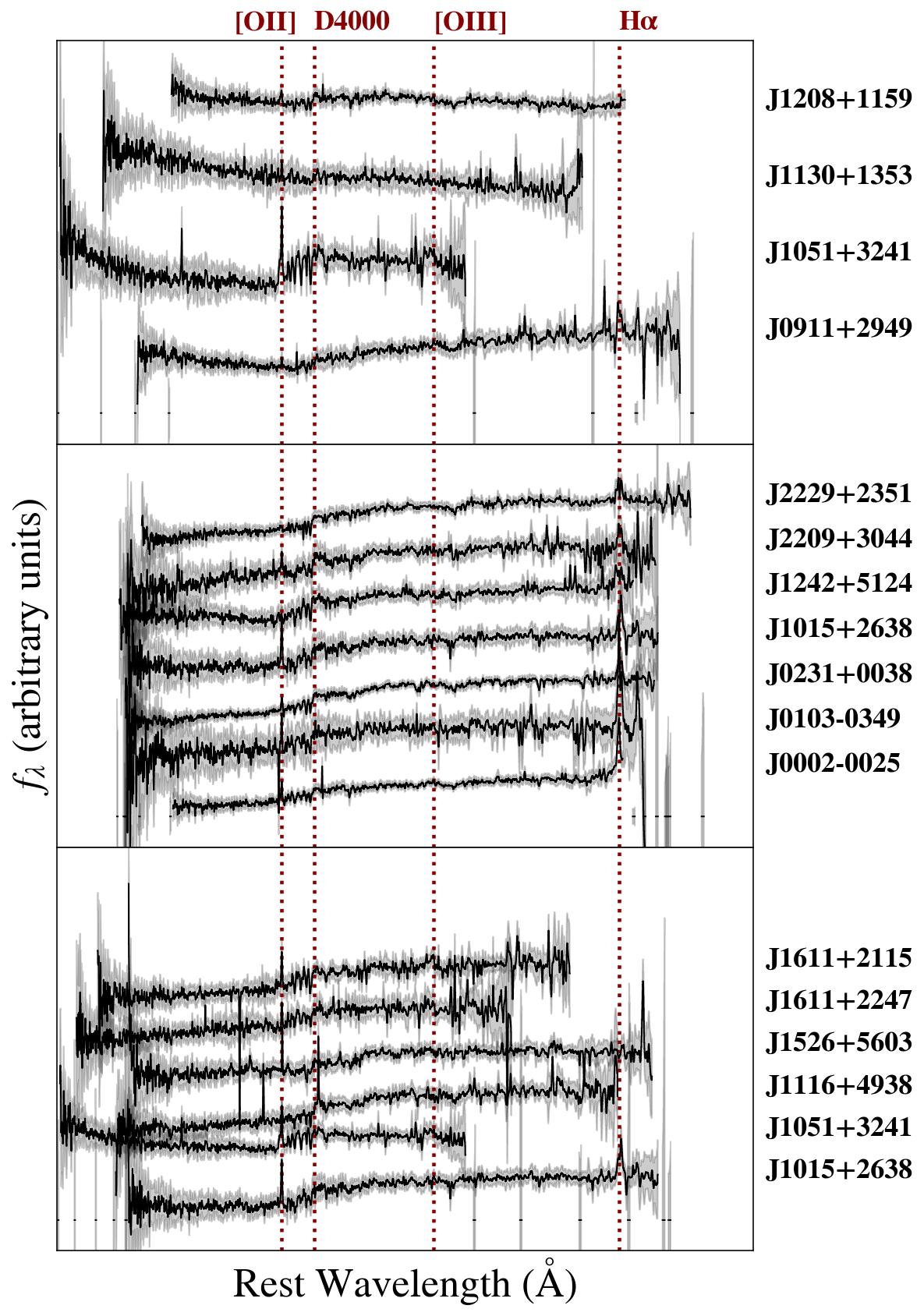}
    \caption[OQQ spectra showing selected subsets.]{OQQ spectra showing selected spectra with the following properties: (top) spectra showing blue rises, (middle) spectra showing strong and/or broad \ha\ emission, (bottom) spectra showing narrow \oii\ emission.}
    \label{fig:oqq_sdss_spectra_details}
\end{figure*}

\begin{itemize}
    \item \textbf{Blue continuum shapes}: although any sources with extremely blue spectra were removed in Section~\ref{sec:sampleselection}, some remain with a blue upturn below the \SI{4000}{\angstrom} break (for example, sources shown in Figure~\ref{fig:oqq_sdss_spectra_details}, top), which may indicate some contribution to the continuum from the accretion disc. This is a small number of sources and are unlikely to represent a major part of the OQQ population, but the combination of continuum shape and lack of emission lines may make these interesting sources. Alternatively, this blue emission could be due to recent star formation, especially if seen in combination with other signatures such as \oii.
    \item \textbf{Broad emission lines}: generally emission lines are low level or narrow for the majority of the sample, but there are some exceptions. J0948+0958 in Figure~\ref{fig:oqq_sdss_spectra_details} (middle) shows broad \ha\ emission, possibly implying a source with greater velocity dispersion than expected from the galaxy, and likely AGN-related. \ha\ is present in several more sources but generally less broad. It is outside of the clean spectroscopic range for a large proportion of the OQQs, making judgement of population properties in this case harder.
    \item \textbf{SF vs. dead galaxy indications}: some SED fitting shows a contribution to MIR luminosities from SF (see Section~\ref{sec:agnfitter}), but only a small fraction of OQQs have sufficient data to draw reliable conclusions about their properties from these fits. Nevertheless, it implies that not all of the OQQs are dead galaxies with very little star formation, and observed properties of some spectra also lead to this conclusion. Narrow emission lines such as \oii\ may indicate a younger stellar population and thus recent or ongoing star formation, and some sources (e.g. sources in Figure \ref{fig:oqq_sdss_spectra_details}, (bottom)) may show this. A strong 4000 \AA\ break and Ca H and K absorption lines could instead imply an older population. However, as discussed in Section \ref{sec:d4000}, this may not be conclusive.
\end{itemize}

We can conclude from inspection of the spectra (along with other properties) that it is unlikely that OQQs represent a \textit{single intrinsic type}. Rather, it is more likely that they represent multiple classes that, due to obscuration or accretion properties, appear similar in terms of OQQ selection (see Section~\ref{sec:nature}). Future work is needed to effectively and conclusively distinguish between intrinsic properties.

\subsection{Continuum SNR check} \label{app:snrcheck}

To eliminate sources where the spectrum in the region of \oiii\ is significantly noisy -- i.e. that there may be \oiii\ present at a low level -- we estimate the noise as follows (summary of results in Table \ref{tab:oiiisummary} as \lq Poor S/N\rq):

\begin{enumerate}
    \item Cut the spectrum to two regions on either side of the expected position of \oiii\ (vertical bands in Fig. \ref{fig:contfitcut}, bottom right panel), avoiding any other potential emission or absorption lines.
    \item Estimate the noise in these regions as the standard deviation on the flux data.
    \item Estimate the signal as the mean on the flux data.
    \item Discard any candidate OQQ with the ratio of these $>$2.
\end{enumerate}

\begin{table}
	\centering
	\caption{Counts of objects removed or kept at all stages of checks outlined in this Appendix.}
	\label{tab:oiiisummary}
	\begin{tabular}{lrlrl}
\hline
                                       &      Original candidates &    \% \\
\hline
                                 Total &                     1520 &        \\
                Lower bound flux $<$ 0 &                      203 &  13.4  \\
\hline
                Keep (after basic vis check) &                 98 &  48.3 \\
                Remove (visible \oiii) &                       48 &  23.6 \\
                        Remove (noisy) &                       42 &  20.7 \\
                  Remove (wrong class) &                       14 &   6.9 \\
\hline
                Keep (after detailed vis check) &              64 &  64.3 \\
                Uncertain redshift &                           20 &  20.4 \\
                Poor S/N &                                     12 &  12.2 \\
                Wrong shape continuum &                          3 &  3.1 \\
\hline
\end{tabular}

\end{table}

\subsection{Duplicate Spectra}

For the fitting and selection process, and the properties in the main sections of the paper, we use only the spectra assigned by \sdss\ as {\tt SciencePrimary} (a flag that indicates that this spectrum is considered the best available for this source). However, for a minority of sources there are multiple spectra available, so we compare these to see if there is any significant difference in flux, class, redshift, or emission line flux.

Most of these duplicate spectra are either well matched, or are candidates that are dismissed for high levels of noise. We discuss the exceptions to this below.

\subsubsection{J0002: 0.592036, -0.4320923}

Two available spectra; generally consistent in appearance and flux, but one has been assigned the \sdss\ subclass `Star-forming' (our selection process requires subclass `null'). This may be due to a change in emission line height. We choose to keep this candidate based on its otherwise promising appearance.

\subsubsection{J0231: 37.99002, 0.648491}

Fifteen available spectra; generally consistent in appearance and flux, but again one has been assigned the \sdss\ subclass `Star-forming'. We also choose to keep this candidate based on its otherwise promising appearance; as the potentially star-forming galaxy is only one of the available spectra, and the rest are `null', as required.

\subsubsection{J0911: 137.96392, 29.824825}

Three available spectra. All with consistent \sdss\ class and redshift, but clear changes in flux between spectra, particularly noting a change in shape in one. The primary spectra is the most recent, so we have retained this object in the sample, but note that it may be variable.

\section{Colour images} \label{app:images}

\begin{figure*}
	\includegraphics[width=\textwidth,keepaspectratio]{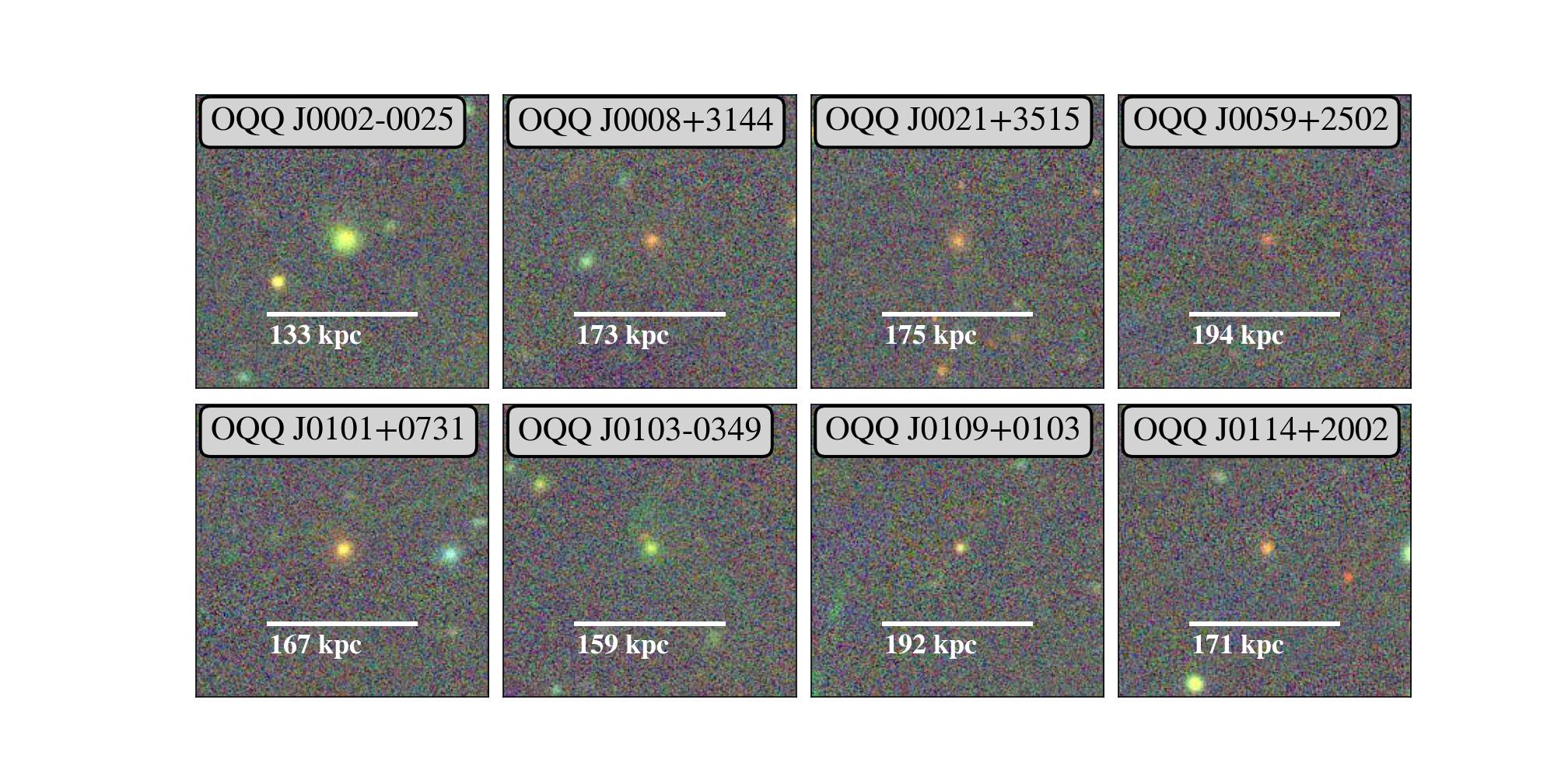}
	\vspace*{-12mm}
    \caption{PanSTARRS colour images of the first 8 OQQs by RA. Bars show approximate scale of objects based on redshift. Full sample available online.}
    \label{fig:panstarrsimages}
\end{figure*}

This section presents PanSTARRS colour images of a sub-sample of OQQs in Figure \ref{fig:panstarrsimages}. The full sample is available online.

\section{X-ray data sources} \label{app:xray_sources}

Here we summarise the source papers used for QSO2 X-ray information and observation IDs of OQQ data used in Section \ref{sec:xrayobs}.

\begin{table*}
	\centering
	\caption{X-ray and \oiii\ data for QSO2s used in Section \ref{sec:xrayobs}. MIR data is from this work, \oiii\ is from \citet{reyes_space_2008} and \citet{yuan_spectroscopic_2016}. X-ray data is indicated in column (8): (a) \citet{vignali_quest_2006}, (b) \citet{vignali_discovery_2010}, (c) \citet{lamastra_bolometric_2009}.}
	\label{tab:qso2_xray}
	\begin{tabular}{lrrrrrrc}
\hline
 \thead{Name\\ \\(1)}      &   \thead{RA (J2000)\\deg.\\(2)} &   \thead{Dec (J2000)\\deg.\\(3)} &   \thead{log \lmir\\erg s$^{-1}$ \\(4)} &   \thead{log $L_{2-10 \textrm{ keV}}$\\erg s$^{-1}$ \\(5)} &   \thead{log $L_{\textrm{\oiii}}$\\L$_\odot$ erg s$^{-1}$ \\(6)} &   \thead{log \nh\\cm$^{-2}$\\(7)} & \thead{Reference\\ \\(8)}   \\
\hline
 WISEA J012341.47+004435.8 &               20.923 &                 0.743 &                                                       44.83 &                                                      44.53 &                                                                    9.13 &                               23.16 & (b)                         \\
 WISEA J080154.26+441234.0 &              120.476 &                44.209 &                                                       45.04 &                                                      44.62 &                                                                    9.58 &                               23.63 & (a)                         \\
 WISEA J081253.10+401859.9 &              123.221 &                40.317 &                                                       45.05 &                                                      44.23 &                                                                    9.39 &                               22.33 & (b)                         \\
 WISEA J083945.98+384318.9 &              129.942 &                38.722 &                                                       44.97 &                                                      44.26 &                                                                    9.71 &                               22.52 & (c)                         \\
 WISEA J115314.38+032658.6 &              178.310  &                 3.450  &                                                       45.04 &                                                      44.08 &                                                                    9.61 &                               22.19 & (a)                         \\
 WISEA J122656.47+013124.3 &              186.735 &                 1.523 &                                                       45.30  &                                                      44.63 &                                                                    9.66 &                               22.42 & (a)                         \\
 WISEA J122845.73+005018.8 &              187.191 &                 0.839 &                                                       45.11 &                                                      43.54 &                                                                    9.28 &                               23.18 & (b)                         \\
 WISEA J164131.72+385840.7 &              250.382 &                38.978 &                                                       45.48 &                                                      44.87 &                                                                    9.92 &                               22.74 & (a)                         \\
\hline
\end{tabular}
\end{table*}

\begin{table*}
	\centering
	\caption{In Section \ref{sec:xrayobs} we examined available serendipitous X-ray observations of OQQs, where available. The targeted observations obtained for J0751 are listed in Section \ref{sec:J0751xray}. The remainder of instruments and observation IDs used are listed here.}
	\label{tab:oqq_xray}
	\begin{tabular}{lll}
\hline
 \thead{Name\\(1)}   & \thead{Instrument\\(2)}   & \thead{Observation ID(s)\\(3)}                                                   \\
\hline
 OQQ J0002-0025      & \xrt                      & 38112001,38112003,38112009,38112002,38112007,38112008,38112005,38112006,38112004 \\
 OQQ J0109+0103      & \xrt                      & 47886001                                                                         \\
 OQQ J0231+0038      & \xmm                      & 652400801                                                                        \\
 OQQ J0751+4028      & \xmm,   \nustar           & 0884080101,60701009002                                                                        \\
 OQQ J1051+3241      & \xmm                      & 781410101,781410201                                                              \\
 OQQ J1611+2247      & \chandra                  & 15679                                                                            \\
 OQQ J1617+0854      & \xrt                      & 46227001,46227006,46227004,46227005,46227003                                     \\
\hline
\end{tabular}

\end{table*}

%%%%%%%%%%%%%%%%%%%%%%%%%%%%%%%%%%%%%%%%%%%%%%%%%%

% Don't change these lines
\bsp	% typesetting comment
\label{lastpage}
\end{document}